\documentclass[twocolumn,trackchanges]{aastex631}
\usepackage{booktabs} 
\usepackage{amsmath} 
\usepackage{soul} 
\usepackage{multirow}
\usepackage{threeparttable}
\usepackage{subfigure}
\usepackage{url}
\usepackage{bm}
\graphicspath{{./}}

\newcommand\appendixfootnote[1]{\footnote{\hsize=\columnwidth\advance\hsize by-18pt\relax#1}}

\begin{document}
    \title{Target of Opportunity Observations Detectability of Kilonovae with WFST}
    \correspondingauthor{Zheng-Yan Liu, Wen Zhao}
    \email{ustclzy@mail.ustc.edu.cn, wzhao7@ustc.edu.cn}

    \author[0000-0002-2242-1514]{Zheng-Yan Liu}
    \affiliation{CAS Key Laboratory for Researches in Galaxies and Cosmology, Department of Astronomy, University of Science and Technology of China, Chinese Academy of Sciences, Hefei, Anhui 230026, China}
    \affiliation{School of Astronomy and Space Sciences, University of Science and Technology of China, Hefei 230026, China}
    
	\author[0000-0003-4959-1625]{Zhe-Yu Lin}
	\affiliation{CAS Key Laboratory for Researches in Galaxies and Cosmology, Department of Astronomy, University of Science and Technology of China, Chinese Academy of Sciences, Hefei, Anhui 230026, China}
	\affiliation{School of Astronomy and Space Sciences, University of Science and Technology of China, Hefei 230026, China}
	
	\author[0000-0001-6319-0866]{Ji-Ming Yu}
	\affiliation{Department of Astronomy, School of Physics and Astronomy, Shanghai Jiao Tong University, Shanghai, 200240, China}

	\author[0000-0001-7081-8623]{Hui-Yu Wang}
	\affiliation{CAS Key Laboratory for Researches in Galaxies and Cosmology, Department of Astronomy, University of Science and Technology of China, Chinese Academy of Sciences, Hefei, Anhui 230026, China}
	\affiliation{School of Astronomy and Space Sciences, University of Science and Technology of China, Hefei 230026, China}
 
    \author{Gibran-Marc Mourani}
    \affiliation{CAS Key Laboratory for Researches in Galaxies and Cosmology, Department of Astronomy, University of Science and Technology of China, Chinese Academy of Sciences, Hefei, Anhui 230026, China}
    \affiliation{School of Astronomy and Space Sciences, University of Science and Technology of China, Hefei 230026, China}
    
    \author[0000-0002-1330-2329]{Wen Zhao}
    \affiliation{CAS Key Laboratory for Researches in Galaxies and Cosmology, Department of Astronomy, University of Science and Technology of China, Chinese Academy of Sciences, Hefei, Anhui 230026, China}
    \affiliation{School of Astronomy and Space Sciences, University of Science and Technology of China, Hefei 230026, China}

    
    \author[0000-0002-7835-8585]{Zi-Gao Dai}
    \affiliation{CAS Key Laboratory for Researches in Galaxies and Cosmology, Department of Astronomy, University of Science and Technology of China, Chinese Academy of Sciences, Hefei, Anhui 230026, China}
    \affiliation{School of Astronomy and Space Sciences, University of Science and Technology of China, Hefei 230026, China}
    
    \begin{abstract}
    	Kilonovae are approximately thermal transients, produced by mergers of binary neutron stars (BNSs) and NS-black hole binaries. As the optical counterpart of the gravitational wave event GW170817, AT2017gfo is the first kilonova detected with smoking-gun evidence. Its observation offers vital information for constraining the Hubble constant, the source of cosmic $r$-process enrichment, and the equation of state of neutron stars. The 2.5-meter Wide-Field Survey Telescope (WFST) operates at six bands (u, g, r, i, z, w), spanning from 320 to 925 nm. It will be completed in the first half of 2023, and with a field-of-view diameter of 3 degrees, aims to detect kilonovae in the near future. In this article, considering the influence of the host galaxies and sky brightness, we generate simulated images to investigate WFST's ability to detect AT2017gfo-like kilonovae. Due to their spectra, host galaxies can significantly impact kilonova detection at a longer wavelength. When kilonovae are at peak luminosity, we find that WFST performs better in the g and r bands and can detect 90\% (50\%) kilonovae at a luminosity distance of 248 Mpc (338 Mpc) with 30 s exposures. Furthermore, to reflect actual efficiency under target-of-opportunity observations, we calculate the total time of follow-up under various localization areas and distances. We find that if the localization areas of most BNS events detected during the fourth observing (O4) run of LIGO and Virgo are hundreds of deg$^2$, WFST is expected to find $\sim$30\% kilonovae in the first two nights during O4 period.
    \end{abstract}
    \section{introduction}\label{sec_intro}
    	Mergers of binary neutron stars (BNSs) and neutron star-black hole (NS-BH) binaries have been thought to generate the neutron-rich ejecta through rapid neutron capture ($r$-process) nucleosynthesis \citep{lattimer1974,lattimer1977,eichler1989}. The expanding ejecta heated up by the radioactive decay of $r$-process nuclei can produce a type of transient whose luminosity is approximately a thousand times brighter than a typical nova theoretically, so named as ``kilonova" \citep{Li_1998,10.1111/j.1365-2966.2010.16864.x}. In addition, it has been proposed that the merger of BNS or NS-BH can also produce short-duration gamma-ray bursts (sGRB) \citep{pacynski1986,narayan1992,popham1999}, thus sGRB emission is expected to accompany a kilonova and a gravitational-wave (GW) burst if the jet is pointing towards Earth \citep{2012ApJ...746...48M,tanvir_kilonova_2013,10.1093/mnras/stz2255,Lamb_2019,jin_kilonova_2020}. 
    	\par
    	\begin{table*}[t]
    		\footnotesize
    		\centering
    		\begin{tabular}{*{13}{c}}
    			\toprule
    			\multirow{1}*{\bf Telescope} &\multicolumn{6}{c}{\multirow{1}*{limiting magnitude \& sky brightness}} & {\bf diameter }   & {\bf readnoise}   & {\bf pixel scale } & {\bf image quality} & {\bf FoV}&{\bf Ref}\\
    			&\multicolumn{6}{c}{(mag \& mag/arcsec${^2}$)}& (m) &  (e/pixel) & (arcsec) & (arcsec) & (deg$^2$)\\
    			\midrule
    			\multirow{3}*{WFST} & u & g & r & i & z & w & \multirow{3}*{2.5}&\multirow{3}*{10}&\multirow{3}*{0.33}&\multirow{3}*{1.0}&\multirow{3}*{6.55}&\multirow{3}*{(1)}\\
    			&22.40&23.35&22.95&22.59&21.64&22.96\\
    			&22.29&22.12&21.58&21.29&20.29&...\\
    			\hline
    			\multirow{3}*{LSST} & u & g & r & i & z & y &\multirow{3}*{8.4}&\multirow{3}*{18}&\multirow{3}*{0.2}&\multirow{3}*{0.80}&\multirow{3}*{9.6}&\multirow{3}*{(2)}\\
    			& 23.87&24.82&24.36&23.93&23.36&22.47\\
    			& 22.96&22.26&21.20&20.48&19.60&18.61\\
    			\hline
    			\multirow{3}*{ZTF} &  \multicolumn{2}{c}{g} & \multicolumn{2}{c}{r}& \multicolumn{2}{c}{i}&\multirow{3}*{1.2}&\multirow{3}*{8}&\multirow{3}*{1.0}&\multirow{3}*{2.0}&\multirow{3}*{47}&\multirow{3}*{(3)} \\
    			&\multicolumn{2}{c}{21.1}&\multicolumn{2}{c}{20.9}&\multicolumn{2}{c}{20.2}\\
    			&\multicolumn{2}{c}{21.8}&\multicolumn{2}{c}{20.7}&\multicolumn{2}{c}{19.9}\\
    			\bottomrule
    		\end{tabular}
    		\caption{The parameters and information of telescopes used in simulation for WFST, ZTF and LSST. For each telescope, the upper and bottom rows are 5-$\sigma$ limiting magnitude for a 30-second exposure and dark night sky brightness individually. For the sky brightness of ZTF, we use the measurement of P200 to represent the sky background of Palomar observatory. Reference: (1) \cite{lin_prospects_2022,lei_limiting_2023}; (2) \cite{ivezic_lsst_2019,LSST:2010:Misc} (3) \cite{bellm_zwicky_2019,P200:2007:Misc}.}
    		\label{tab:teles}
    	\end{table*}
    	On 2017 August 17 at 12:41:04.47 UTC, the first BNS merger GW source GW170817 was detected by the Advanced Laser Interferometer Gravitational-Wave Observatory (LIGO) and Virgo with a false-alarm-rate estimate of less than one per $8.0 \times 10^4$ years \citep{abbott_gw170817_2017}. After the GW detection alert was triggered ($\sim1.7$ s), the Fermi Gamma-ray Space Telescope and the International Gamma-Ray Astrophysics Laboratory (INTEGRAL) detected the sGRB GRB170817A which lasted about 2 s \citep{goldstein_ordinary_2017,savchenko_integral_2017}. Based on the sky map constrained by the GW and sGRB signal, telescopes all around the world started to search for the optical counterpart of GW170817. The Swope team was the first to observe the candidate, declaring it to be located in NGC4993, which was subsequently confirmed to be a kilonova by other telescopes' observations, and named as AT2017gfo.
    	 \citep[e.g.,][]{coulter_swope_2017,pian_spectroscopic_2017,tanvir_emergence_2017,cowperthwaite_electromagnetic_2017,kasliwal_illuminating_2017,shappee_early_2017,evans_swift_2017,smartt_kilonova_2017}. 
    	\par
    	A telescope with a large field-of-view (FoV) is necessary for searching for the electromagnetic counterpart of the merger, due to the large localization area with $10^1$--$10^3$ ${\deg}^2$ from GW detectors. The 2.5-meter Wide-Field Survey Telescope (WFST) 
    	will be installed at the summit of the Saishiteng Mountain near Lenghu which is planned to be completed in the first half of 2023, and commence operations subsequently. With a FoV of 6.55 ${\deg}^2$, WFST will scan the northern sky in six optical bands (u, g, r, i, z, w). Reaching a depth of 22.40, 23.35, 22.95, 22.59, 21.64, 22.96 AB mag in a nominal 30-second exposure in the optical bands respectively, which meets the scientific requirements for kilonova detection. There is a median seeing of 0.75 \arcsec{} \citep{deng_lenghu_2021} in the Lenghu site, which provides an ideal environment for WFST. Given these parameters and its geographical location, WFST can bridge the longitudinal gap between other wide-field instruments and has the potential to be one of the major GW follow-up instruments in the northern hemisphere for the upcoming fourth (O4) and fifth observing (O5) run of ground-based GW detectors.
    	\par
		In the third observing (O3) run of LIGO and Virgo, in order to search for new kilonovae, many instruments including the Zwicky Transient Facility \citep[ZTF;][]{bellm_zwicky_2019,graham_zwicky_2019} and the Dark Energy Camera \citep[DECam;][]{flaugher_dark_2015} have made extensive follow-up observations. However, there was no conclusive evidence for kilonovae detection by the end of the O3 run. \citep{andreoni_growth_2019,Andreoni_2020,ackley_observational_2020,kasliwal_kilonova_2020,anand_optical_2021,tucker_soargoodman_2022}. To improve follow-up observations with WFST in upcoming O4 and O5, we need to handle the trade-off between exposure time and sky coverage, using previous observations for reference. Recently, some studies on kilonova detectability predicted the expected number of kilonovae for various wide-field instruments \citep{cowperthwaite_comprehensive_2015,cowperthwaite_lsst_2019,scolnic_how_2017,zhu_kilonova_2021,chase_kilonova_2022,wanghuiyu_2022}. Based on the different observation cadences, \cite{cowperthwaite_lsst_2019} explored the efficiency of searching for kilonovae randomly with the Large Synoptic Survey Telescope \citep[LSST;][]{lsst_science_collaboration_science-driven_2017,ivezic_lsst_2019} in the Vera C. Rubin Observatory. The research concluded that it is more effective to search for kilonovae after a GW trigger than serendipitous observations. By considering the afterglow of sGRB and the angle of view, for AT2017gfo-like kilonovae, \cite{zhu_kilonova_2021} studied the prospect of finding different combinations of kilonova and GRB afterglow. \cite{chase_kilonova_2022} calculated the detection depth of thirteen wide-field instruments based on a specific kilonova model and a grid of parameters. In these works, the influence of image subtraction and the host galaxy on kilonova detection is always ignored, therefore the detection depth will be overestimated, resulting in more kilonovae predicted to be found. In practice, evaluating these effects is also needed to select the appropriate exposure time in a target-of-opportunity (ToO) observation.
    	\par
\begin{figure}[h]
	    	\centering
	    	\includegraphics[scale=0.43]{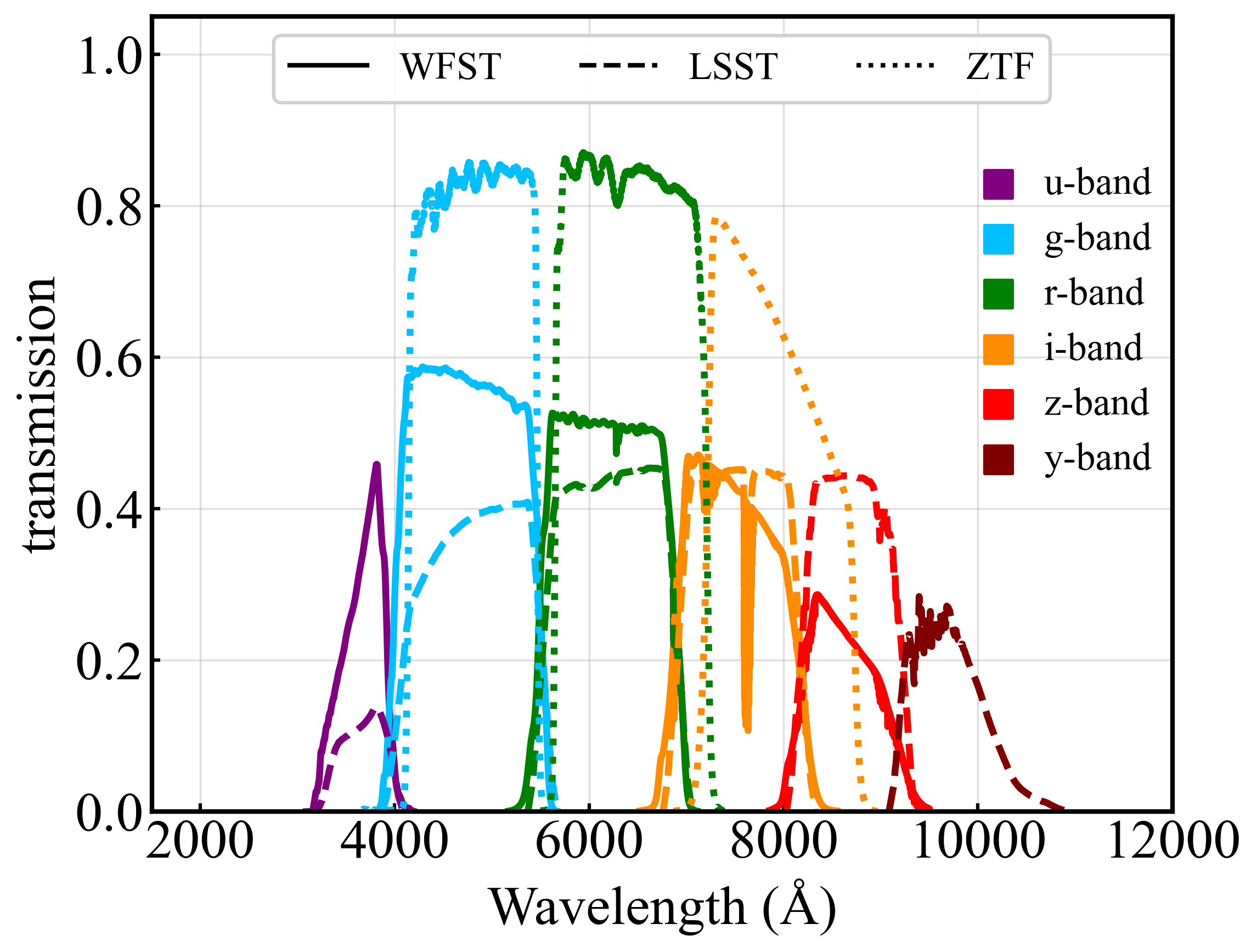}
	    	\caption{The transmission of the telescope system of WFST, ZTF and LSST.}
	    	\label{fig:transmission}
	    \end{figure}

        In this work, we assess the ability of WFST to detect AT2017gfo-like kilonovae with mock observations. We focus on five of WFST's optical band filters, excluding the w-band, which is a broad bandpass and is not helpful to measure color information of kilonovae (The transmission of other bands is shown in Figure \ref{fig:transmission}). By considering galaxies in a synthetic galaxy catalog as hosts of kilonovae, we investigate and quantify the influence of image subtraction and the host galaxy on kilonova searches. The methods of mock observation and its verification and accuracy are described in Section \ref{sec_1}. In Section \ref{sec_2}, the details about adding a host galaxy into an image are introduced, and we display and discuss the simulation result of kilonova detectability for WFST, LSST and ZTF. In Section \ref{sec_3}, we optimize the exposure time selection upon these results and explore the follow-up capacity of WFST by estimating the average total time spent in ToO observation. Finally, as a complement to the case of AT2017gfo-like kilonova, we discuss the detection ability for other kilonovae and summarize our results in Section \ref{sec_4}. Throughout this study, we adopt a standard $\Lambda$CDM cosmology with parameters $H_0=69.3 \, \text{km s}^{-1}\,\text{Mpc}^{-1}$, $\Omega_M=0.287$ and $\Omega_\Lambda=0.713$ \citep{Hinshaw_2013}.
    \section{simulation process} \label{sec_1}
    	In this section, we introduce the method of simulated image synthesis and test it by computing the kilonova detection depth of WFST with a flat background. The main steps of the synthesis process are as follows:
    	\par
    	\textbf{Kilonova template: }The luminosity evolution of AT2017gfo can be explained and fitted by many kilonova models \citep[e.g.,][]{kasen_origin_2017,Villar_2017,yu_long-lived_2018,bulla_possis_2019,hotokezaka_radioactive_2020,wollaeger_broad_2021}. The main differences between these models are: the mechanism that powers the early emission, the ejecta matter composition and morphology or the calculation of radiation transfer. Given that there are few observations of kilonova except AT2017gfo, these models cannot be distinguished based on the limited observation data. The chance of detecting a kilonova depends on the template which can be chosen from data or be generated by a theoretical model. Considering the lack of constraints to the kilonova models and uncertainty of the model parameters, we choose AT2017gfo as a source of reference in study and obtain its lightcurve by \texttt{MOSFiT}\footnote{\href{https://github.com/guillochon/mosfit}{https://github.com/guillochon/mosfit}}. Using AT2017gfo as the kilonova template leaves out the potential for kilonova diversity, which can affect the overall estimation of kilonova detection. Combined with a certain kilonova model, this influence is discussed specifically in Section \ref{other kilnovae}.
		\par
		\texttt{MOSFiT} is a python package that collects transient templates (e.g., kilonova, tidal disruption event (TDE) and supernova (SN)), which can be used to fit the observations and generate theoretical lightcurves \citep{Guillochon_2018}. The kilonova template in \texttt{MOSFiT} originates from \cite{Villar_2017} where the authors constructed a spherically symmetric model composed of two or three components ejecta matter. We choose the three-component model where these components have different opacities ($\kappa$) fixed in the model, which are named as blue ($\kappa=0.5 \text{ cm}^2 \text{g}^{-1}$), purple ($\kappa=3 \text{ cm}^2 \text{g}^{-1}$) and red ($\kappa=10 \text{ cm}^2 \text{g}^{-1}$) components. There are three more free parameters to describe each component: mass ($M_{\text{ej}}$), velocity ($v_{\text{ej}}$) and temperature floor ($T_c$). In addition, the authors introduced a variance parameter ($\sigma$) in the likelihood function, which encompasses additional uncertainty in the model and/or data. To generate the lightcurve, the parameters of best-fit result in \cite{Villar_2017} are adopted: $M_{\text{ej}}^{\text{blue}}=0.020 M_{\odot}$, $v_{\text{ej}}^{\text{blue}}=0.266c$, $T^{\text{blue}}=674$ K, $M_{\text{ej}}^{\text{purple}}=0.047 M_{\odot}$, $v_{\text{ej}}^{\text{purple}}=0.152c$, $T^{\text{purple}}=1308$ K, $M_{\text{ej}}^{\text{red}}=0.011 M_{\odot}$, $v_{\text{ej}}^{\text{red}}=0.137c$, $T^{\text{red}}=3745$ K, and $\sigma=0.242$. It is worth noting that the lightcurve fitting generated by different kilonova models can be somewhat different \citep{arcavi_first_2018}. Especially for early time ($\delta t < 10\;\text{hr}$) of kilonova emission, the lightcurve mainly depends on the model used due to the lack of observation data \citep{arcavi_optical_2017}.
    	\par
    	\textbf{Image generation: }We generate the simulated images by \texttt{GalSim}\footnote{\href{https://github.com/GalSim-developers/GalSim}{https://github.com/GalSim-developers/GalSim}} which is an open-source project providing a software library for simulating images of astronomical objects such as stars and galaxies in a variety of ways \citep{ROWE2015121}. Taking into account the generation speed and computing resources, we produce simulated images with at least 200$\times$200 pixels and place the target in the center. When a kilonova and its background are added into images, using point-spread functions (PSF) integrated into \texttt{GalSim}, we consider the extension of point-source caused by the optical system and atmospheric seeing. In addition, Poisson noise and readout noise can also be added to images by the built-in noise generators. The image quality and readout noise we use in the simulation are shown in Table \ref{tab:teles}.
    	\par
    	\textbf{Photometry and evaluation criteria: }By subtracting simulated reference images from science images, we can obtain difference images and make photometry with \texttt{PythonPhot}\footnote{\href{https://github.com/djones1040/PythonPhot}{https://github.com/djones1040/PythonPhot}} to judge whether the kilonova can be detected or not. \texttt{PythonPhot} is a python package translated from the DAOPHOT-type photometry procedures of the IDL AstroLib photometry algorithms \citep{2015ascl.soft01010J,landsman1993idl}. When a kilonova is relatively faint compared to its background, the measurement error of the PSF-fitting photometry tool of \texttt{PythonPhot} may be overestimated. So we use aperture photometry to assist PSF-fitting photometry to get the SNR of the source. The diameter of the aperture is taken as 1.5 times the FWHM in each band. The FWHM of each band is computed based on the image quality parameters listed in Table \ref{tab:teles}. Among all the results presented in this article, we set the threshold of SNR as 5 to judge whether kilonovae can be detected or not.
    	\par
    	\begin{figure}[htbp]
    		\centering
    		{\includegraphics[scale=0.45]{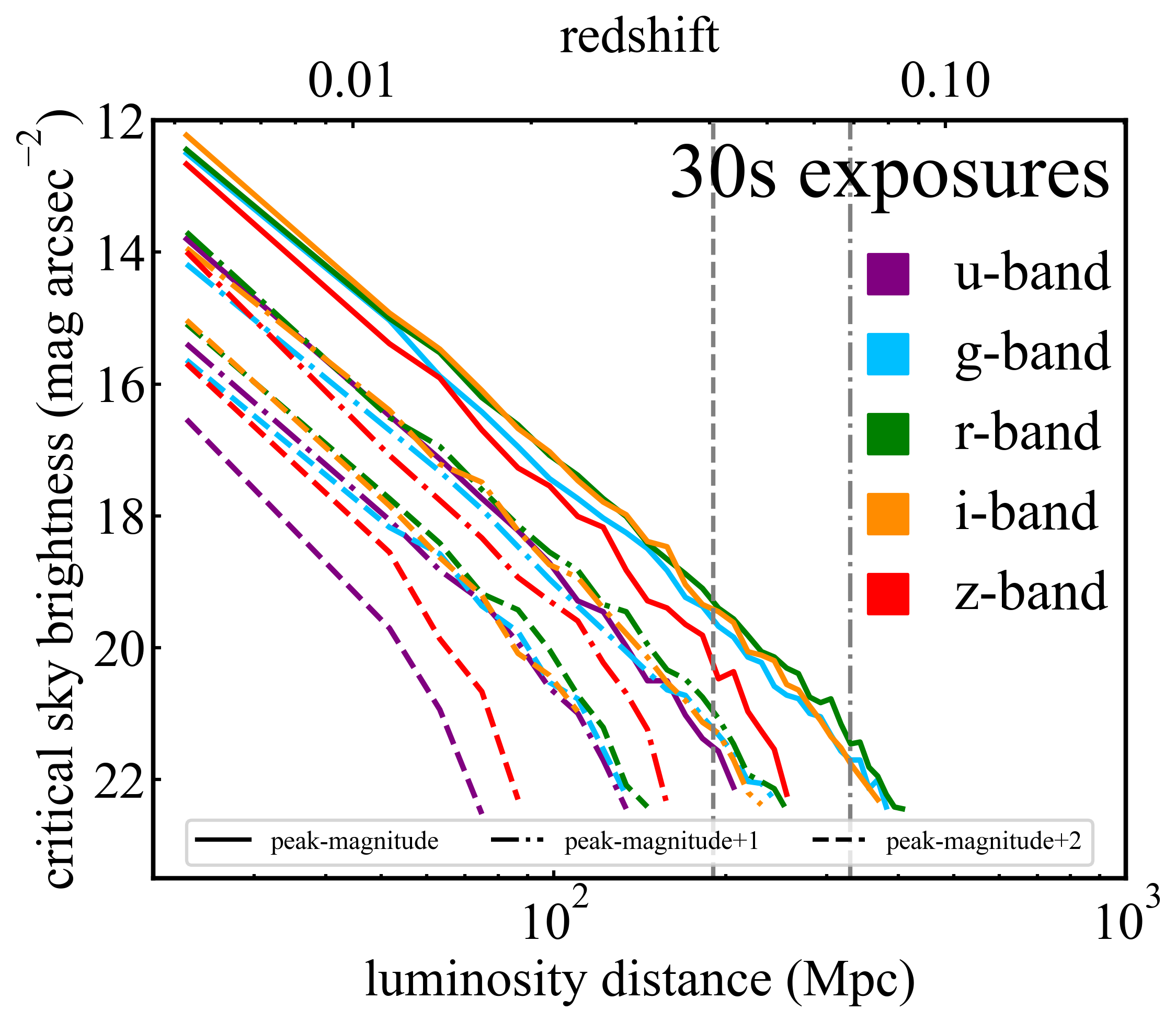} }
    		{\includegraphics[scale=0.45]{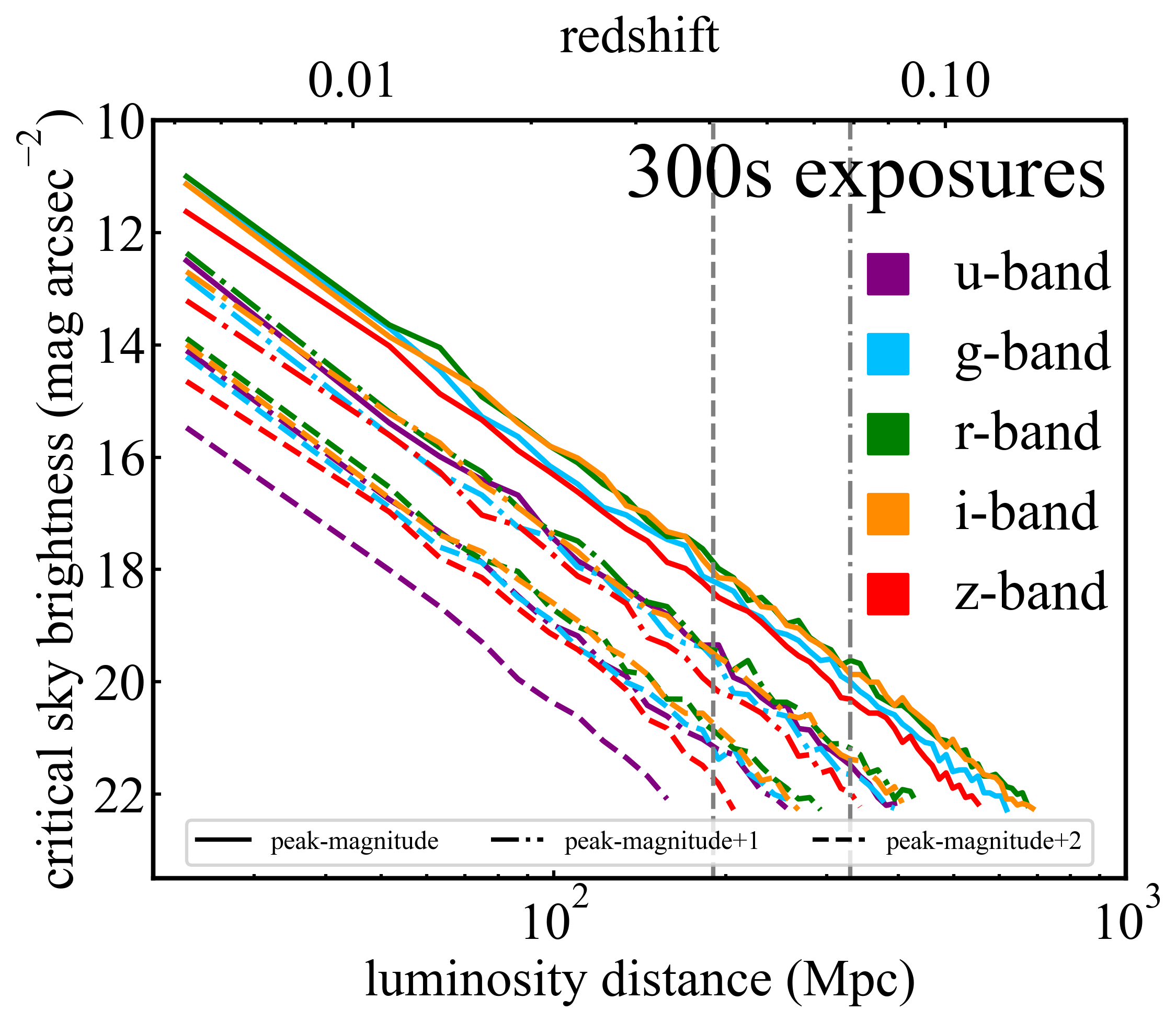} }
    		\caption{At different times of evolution of AT2017gfo-like kilonova, the relationship between luminosity distance and critical sky brightness from image simulation with 30 and 300 s exposures. The grey dashed and dot-dashed lines represent the detection range for BNS of O4 and O5, respectively \citep{abbott_prospects_2020}.}
    		\label{fig:background}
    	\end{figure}
        \begin{table}[htbp]
    	\centering
	    	\begin{tabular}{*{7}{c}}
	    		\toprule
	    		\multirow{2}*{\bf exposure (s)} & \multirow{2}*{\bf luminosity} & \multicolumn{5}{c}{\bf detection depth (Mpc)}\\
	    		\cline{3-7}	& &\bf u &\bf g &\bf r &\bf i &\bf z\\
	    		\midrule
	    		\multirow{3}*{30}&peak&210&357&338&302&202\\
	    		&peak+1 mag&131&233&209&193&123\\
	    		&peak+2 mag&73&131&127&112&69\\
	    		\hline
	    		\multirow{3}*{300}&peak&395&609&562&515&320\\
	    		&peak+1 mag&257&375&366&317&206\\
	    		&peak+2 mag&163&256&234&214&137\\
	    		\bottomrule
	    	\end{tabular}
    	\caption{The detection depths at different times of evolution of AT2017gfo-like kilonova for WFST.} 
    	\label{tab:depth}
   		\end{table}
    	To test and verify the ability of the simulation process, we consider a simple case where we insert kilonovae into the images' center and choose a uniform background of sky brightness to compute its influence on detection. For kilonovae at different distances, we set a range of sky brightness to obtain the critical value at which the telescope can detect a kilonova. The relationships between luminosity distance and critical sky brightness in each bandpass of the 30 s and 300 s exposures are shown in Figure \ref{fig:background}. In this simulation, for each band we choose the peak luminosity of the kilonova as well as another two cases: one and two magnitudes fainter than the peak. The comparison of these results can reflect the change of detection depth as the kilonova evolves. From the observation of AT2017gfo, its magnitude faded at $\sim 1$ mag per day in the g and r bands \citep{kasliwal_kilonova_2020}, thus the other two cases (one and two magnitudes fainter than the peak) roughly correspond the 2nd or 3rd day after a BNS merger. We find that g, r and i bands are better choices for WFST to search for kilonova because of their deeper detection depth for a given exposure time. 
    	\par
    	According to the observation condition of the Lenghu site reported in \cite{deng_lenghu_2021}, measured in the bandpass from 400 nm to 600 nm, during a fully clear new Moon phase, the night-sky brightness is 22.3 mag arcsec$^{-2}$ in V-band at most. The average night-sky brightness is around 22.0 mag arcsec$^{-2}$ when the Moon is below the horizon. Considering a background mostly due to night sky brightness, which corresponds to a case where the flux from a faint host galaxy can be ignored, we compute the detection depth of WFST. Each band brightness used is shown in Table \ref{tab:teles} and is set the same as that in WFST Science Collaboration (2023). The detection depths of different situations are shown in Table \ref{tab:depth}. Due to the low transmission shown in Figure \ref{fig:transmission} or the effects of higher sky brightness, the detection ranges of u and z bands are relatively low, thus they are not suitable for early searching of the kilonova. To verify the reliability of the result, we calculate the detection depth by comparing the kilonova magnitude and the limiting magnitudes of WFST. Using the limiting magnitudes in Table \ref{tab:teles}, with 30 s exposures, g and i bands have detection depths of 458 and 419 Mpc at luminosity peak. The detection depths in Table \ref{tab:depth} are shallower than them estimated simply using the limiting magnitude for the reason that in a more realistic situation, there will be contribution from noise introduced by image subtraction. Therefore, calculating detection depth with only limiting magnitude can overestimate the detection rate of the kilonova. 
    	\par
   		\section{The influence from host galaxy} \label{sec_2}
	    	Based on the image simulation process introduced in Section \ref{sec_1}, in this section, we insert a host galaxy into the simulated image to consider its influence on the detection of the kilonova.
	    	\par
		    \subsection{Galaxy catalog and building sample}
		      In order to take the host galaxy into account, a specific galaxy catalog needs to be input into the process of image synthesis. For optimizing the follow-up observation of GW alerts of compact binary mergers, some works have assembled known galaxy catalogs into larger, homogenized collections \citep{white_list_2011,dalya_glade_2018,dalya_glade_2021}. Using the observed galaxy catalogs can be more instructive and meaningful in making a follow-up strategy. However, the observed galaxy catalogs are always incomplete at a large distance, and these catalogs do not always have enough information to insert galaxies into simulated images because of the lack of morphological descriptions. So we choose the synthetic galaxy catalog cosmoDC2\footnote{\href{https://github.com/LSSTDESC/cosmodc2}{https://github.com/LSSTDESC/cosmodc2}} to consider the influence of the host galaxy. 
		      \par
		      CosmoDC2 is a large synthetic galaxy catalog which is made for dark energy science with LSST, based on a trillion-particle, $(4.255\;\text{Gpc})^3$ box cosmological \emph{N}-body simulation \citep{korytov_cosmodc2_2019}. CosmoDC2 covers 440 deg$^2$ sky area to a redshift $z=3$, and there are many parameters in this catalog to describe and quantify the properties of each galaxy, such as stellar mass, morphology and spectral energy distribution. These parameters can match well with the interfaces and functions of \texttt{GalSim}. Unlike the observed galaxy catalogs, we can easily add galaxies into images as we know their size and morphology. Nevertheless, bias from the actual case may exist. To make sure that the galaxies in the catalog can be consistent with a real situation, some comparisons including redshift distribution and color distribution have been done and shown in \cite{korytov_cosmodc2_2019}. Hence, choosing galaxies from cosmoDC2 in simulation should fairly reflect the influence of the host galaxy of the real universe.
		      \par      
		      \begin{figure}[h]
			        	\centering
			        	\includegraphics[scale=0.53]{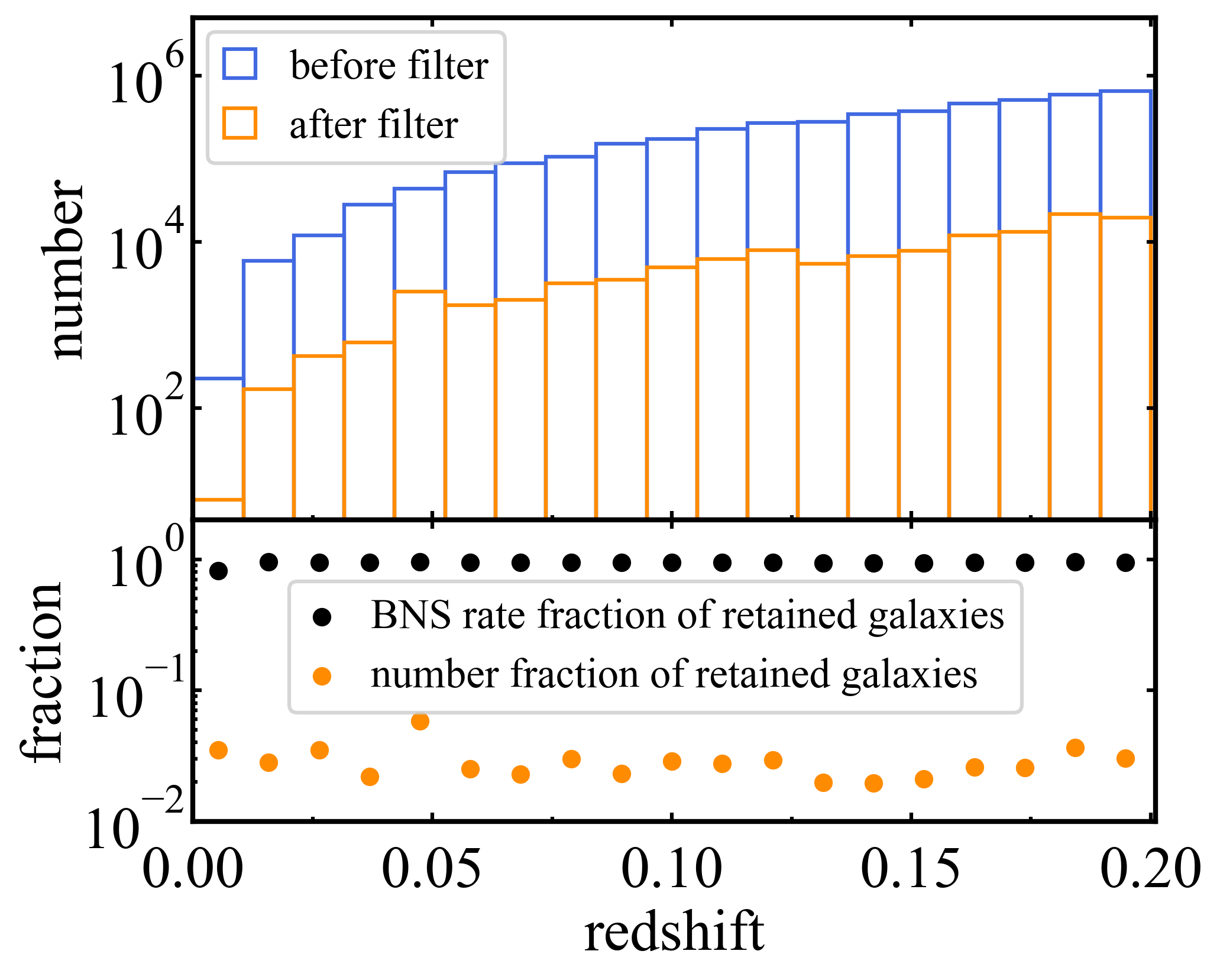}
			        	\caption{At each redshift, the number of galaxies before and after filters and the fraction of the size and BNS merger rate of galaxy sample after filters.}
			        	\label{fig:fraction}
		      \end{figure} 
	    	  \begin{table*}[t]
		    		\centering
		    		\footnotesize
		    		\begin{tabular}{llcc*{3}{r@{\extracolsep{0.2em}}c@{\extracolsep{0.2em}}l}c}
		    			\toprule
		    			\multicolumn{1}{c}{\multirow{2}*{\bf sGRB}} & \multicolumn{1}{c}{\multirow{2}*{\bf Redshift}} &\multirow{2}*{\bf Instrument}&\multirow{2}*{\bf Filter}& \multicolumn{3}{c}{\bf Offset} & \multicolumn{3}{c}{\bf Offset} & \multicolumn{3}{c}{\bf Offset} &\multirow{2}*{\bf Reference}\\ 
		    			&&&&\multicolumn{3}{c}{$(\prime \prime)$} &\multicolumn{3}{c}{(kpc)}&\multicolumn{3}{c}{($r_e$)}&\\
		    			\midrule
		    			\multirow{2}*{150101B}&\multirow{2}*{0.1343}&ACS&F606W&3.07&$\pm$&0.03&7.35&$\pm$&0.07&0.77&$\pm$&0.03&\multirow{2}*{\cite{fong_afterglow_2016}}\\
		    			&&WFC3&F160W&3.07&$\pm$&0.03&7.35&$\pm$&0.07&1.01&$\pm$&0.03&\\
		    			\specialrule{0em}{2pt}{2pt}
		    			160821B&0.162&ACS&F606W&\multicolumn{3}{c}{5.7}&16.40&$\pm$&0.12&\multicolumn{3}{c}{...}&\cite{troja_afterglow_2019}\\
		    			\specialrule{0em}{2pt}{2pt}
		    			\multirow{2}*{170817A}&\multirow{2}*{0.0973}&ACS&F475W, F625W, F775W&10.315&$\pm$&0.007&2.125&$\pm$&0.001&0.64&$\pm$&0.03$^a$&\multirow{2}*{\cite{blanchard_electromagnetic_2017}}\\
		    			&&WFC3&F110W, F160W&10.317&$\pm$&0.005&2.125&$\pm$&0.001&0.57&$\pm$&0.05$^a$&\\
		    			\specialrule{0em}{2pt}{2pt}
		    			181123B&1.754&Gemini-N&$i$&0.59&$\pm$&0.16&5.08&$\pm$&1.38&\multicolumn{3}{c}{...}&\cite{paterson_discovery_2020}\\
		    			\specialrule{0em}{2pt}{2pt}
		    			\multirow{2}*{200522A}&\multirow{2}*{0.5536}&WFC3&F125W&0.155&$\pm$&0.054&1.01&$\pm$&0.35&0.24&$\pm$&0.04&\multirow{2}*{\cite{fong_broadband_2021}}\\
		    			&&WFC3&F160W&0.143&$\pm$&0.029&0.93&$\pm$&0.19&0.24&$\pm$&0.04&\\
		    			\bottomrule
		    		\end{tabular}
		    		\begin{threeparttable}
		    			\begin{tablenotes}
		    				\item[a]{The half-light radius is obtained by the averages of the values from the optical and NIR HST observations in several filters.}
		    			\end{tablenotes}
		    		\end{threeparttable}
		    		\caption{The information of additional measurements of projected offset from 5 sGRBs. The offsets are described in three ways which are angular distance in image, projected distance in kpc and normalized distance by effective radius of galaxy.}
		    		\label{tab:offset}
	    	\end{table*}
	    	\begin{figure*}[t]
		    		\centering
		    		\includegraphics[scale=0.51]{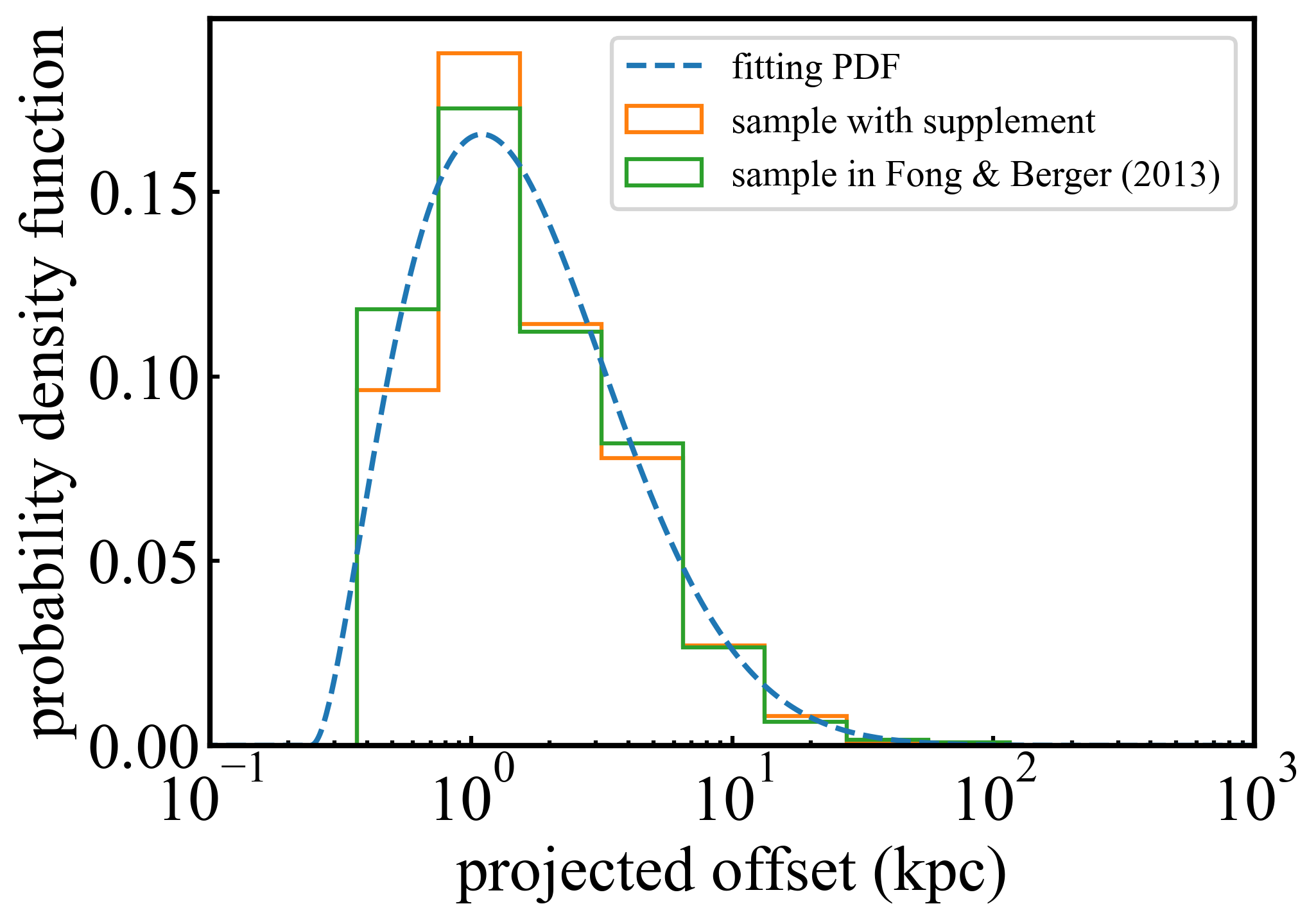}
		    		\includegraphics[scale=0.51]{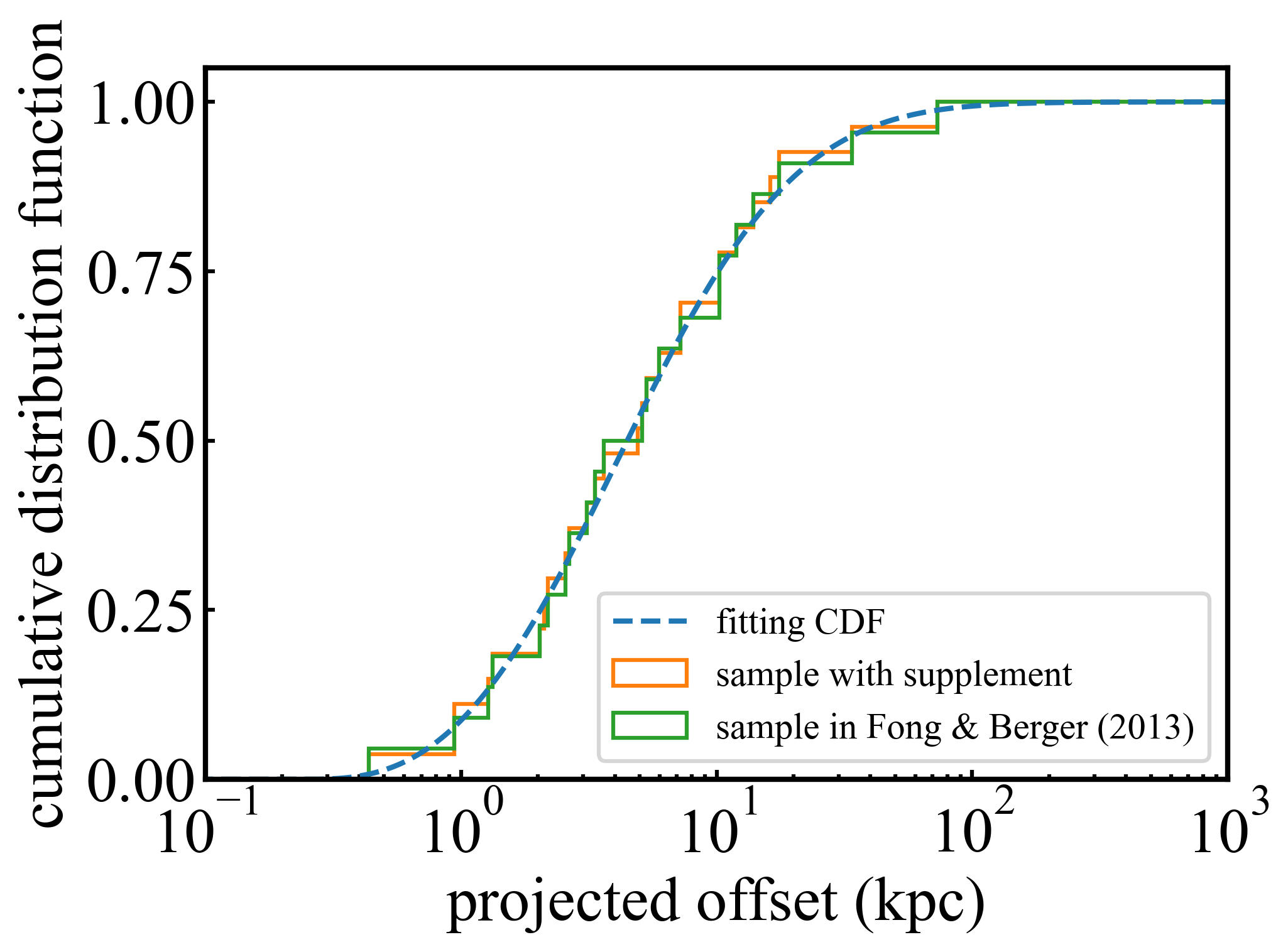}
		    		\caption{The probability density and cumulative distribution of projected offset between the location of sGRB and the center of host galaxy for different samples and their fitting result.}
		    		\label{fig:offset}
			\end{figure*}
			\begin{figure*}
			\centering
			\subfigure[]{	
				\label{subfig:example1}
				\includegraphics[scale=0.3]{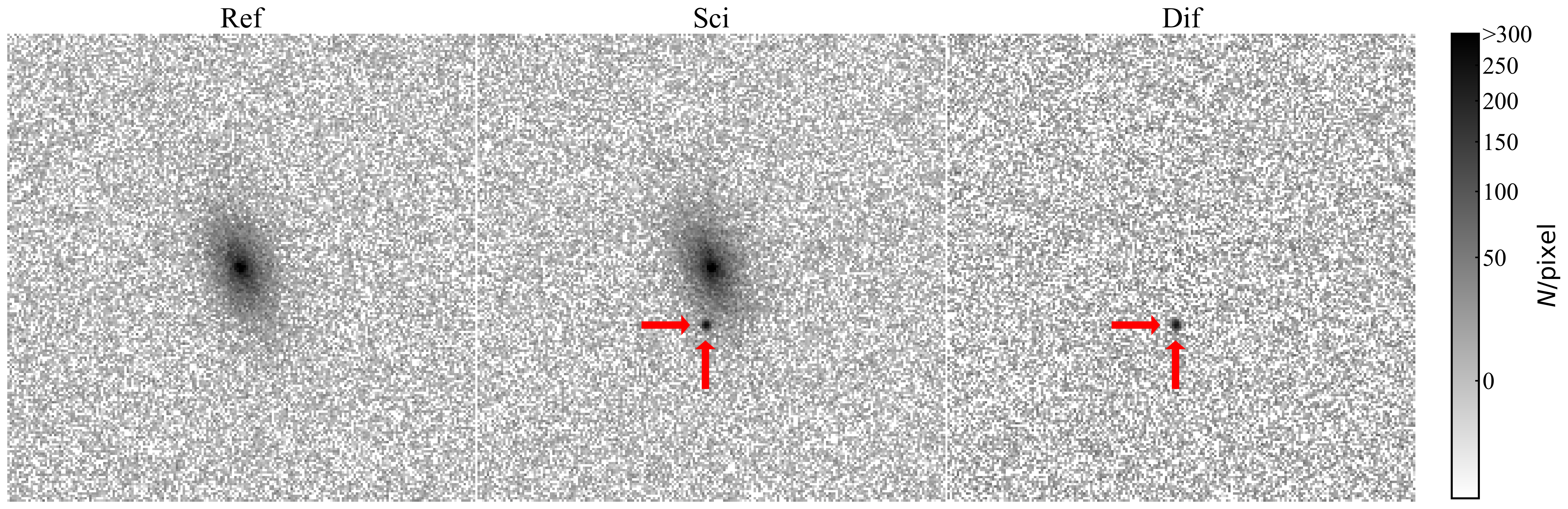}}
			\subfigure[]{	
				\label{subfig:example2}
				\includegraphics[scale=0.3]{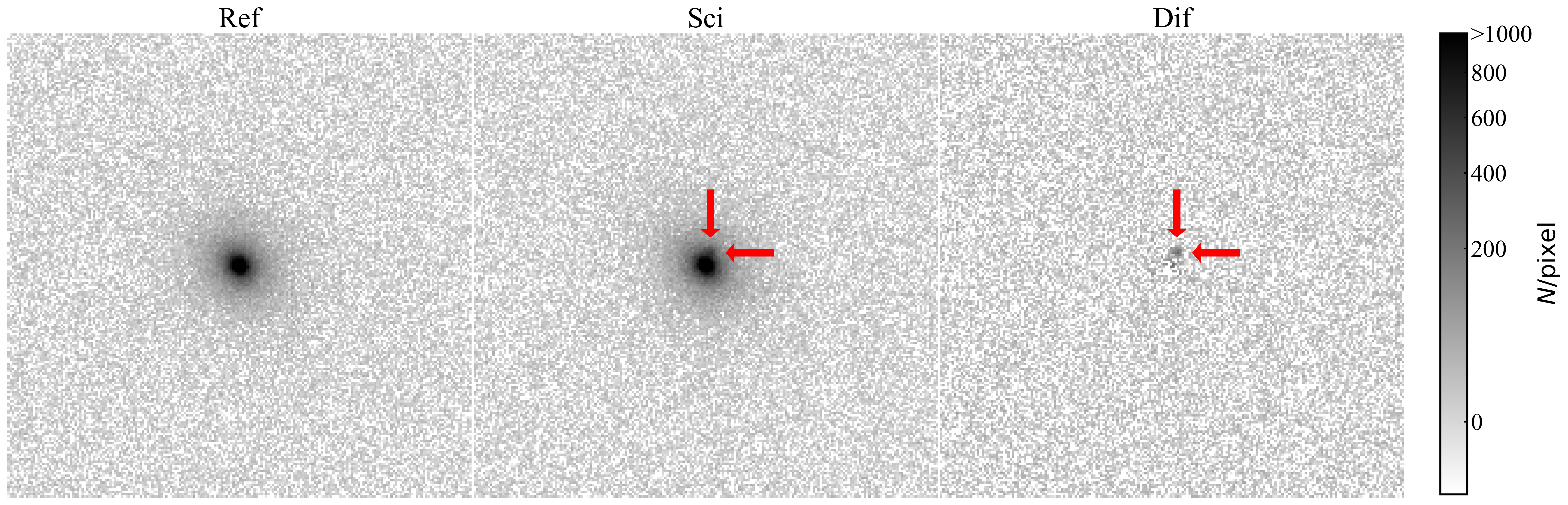}}
			\caption{An example of reference, science and difference images with host galaxies in simulation at redshift $\sim0.056$, \subref{subfig:example1} and \subref{subfig:example2} correspond to cases far from (with a physical offset of $r=1.763$ kpc) and close to (with $r=0.414$ kpc) the host galaxy's nucleus, respectively. The red arrows indicate the location of the kilonovae.}
			\label{fig:example}
			\end{figure*}
			As we wish to consider the hosts of detectable kilonovae, some selection and filters are needed to build the galaxy sample of simulation. According to the sensitivity of the GW detectors for a BNS system of $1.4M_{\odot}+1.4 M_{\odot}$ in O4 and O5 \citep{abbott_prospects_2020}, the typical ranges of detected BNS can reach 190 Mpc and 330 Mpc, which correspond to redshift $z = 0.043 \text{ and } 0.072$, respectively. Given the luminosity of kilonovae, we select galaxies with $z<0.2$ in cosmoDC2 to simulate. Combined with some mechanisms of compact binary formation and the method of population synthesis, the property of host galaxies of merging compact objects can be explored. Several works have attempted to study their properties with this approach \citep{oshaughnessy_effects_2017,cao_host_2018,lamberts_predicting_2018,giacobbo_progenitors_2018,mapelli_host_2018,toffano_host_2019,artale_host_2019,artale_mass_2020}. These works have found that the merger rate of compact binary depends on stellar mass, metallicity and galaxy host type. Considering that the strongest correlation is between the merger rate and the stellar mass of the host galaxy than other parameters \citep{artale_host_2019,artale_mass_2020}, to assign a weight to each galaxy, we adopt the 1D relationship of $z = 0.1$ of the fitting results between stellar mass $(M_*)$ and BNS merger rate $(n_\text{BNS})$ according to \cite{artale_mass_2020}:
			\begin{equation} \label{equation:BNS_rate}
				\begin{aligned}
				\log_{10}(n_\text{BNS}/\text{Gyr})=&(1.038 \pm 0.001)\log_{10}(M_*/M_\odot)\\
				&-(6.090 \pm 0.010).
				\end{aligned}
			\end{equation}
			Comparing the fitting results between $z = 0.1$ and $z = 1$ in \cite{artale_mass_2020}:
			\begin{equation}
			\begin{aligned}
					\log_{10}(n_\text{BNS}/\text{Gyr})=&(1.109 \pm 0.001)\log_{10}(M_*/M_\odot)\\
					&-(6.214 \pm 0.006),
			\end{aligned}
			\end{equation}
			the fitting factors in these two cases are not much different, therefore Eq (\ref{equation:BNS_rate}) can be used under the redshift range of $z \leq 0.2$ in our simulation. We calculate the merger rates of galaxies that satisfy $M_*>{10}^7 M_\odot$. Since the size of the galaxy sample with $z<0.2$ and $M_*>{10}^7 M_\odot$ is also too large, we further narrow down the sample by retaining only those objects with $\log_{10}(n_\text{BNS}/\text{Gyr})>4$, which roughly corresponds to $M_*>{10}^{10} M_\odot$.  After filtering and dividing the galaxies into a series of subsamples with different redshifts, the number of galaxies and BNS merger rate are shown in Figure \ref{fig:fraction}. With the filter of $\log_{10}(n_\text{BNS}/\text{Gyr})>4$, for each galaxy subsample the size is greatly reduced, while the total BNS merger rate changes slightly, which means that most galaxies that contribute little to the merger rate are excluded.
	        \par
		\subsection{Kilonova offset to galaxy center}
			Adding a host galaxy into an image means considering an uneven background with a specific profile, therefore the relative location of the merger to its host is essential to investigate the influence of the host galaxy. There are two ways to explore the offset between the compact binary merger and the galaxy's center: The first is the population synthesis method, and the other is tracing the BNS merger by sGRB.
			\par
			\begin{table*}[ht]
						\centering
						\footnotesize
						\begin{tabular}{c@{\extracolsep{0.4em}}c*{5}c}
							\toprule
							\multirow{2}{*}{\bf exposure time (s)} & \multirow{2}{*}{\bf Efficiency} & \multicolumn{5}{c}{ $\bm{z_{\rm max}\;(D_{\rm L},_{\rm max}}$ \bf Mpc)}\\
							\cline{3-7}&&\bf u&\bf g&\bf r&\bf i&\bf z\\
							\midrule
							\multirow{2}*{30}&90\%&0.039 (175.3)&0.055 (246.7)&0.046 (205.8)&0.038 (168.4)&0.036 (158.5)\\
							&50\%&0.051 (230.4)&0.084 (386.7)&0.078 (357.3)&0.069 (316.0)&0.045 (202.0)\\
							\hline
							\multirow{2}*{90}&90\%&0.06 (270.2)&0.074 (336.8)&0.06 (271.9)&0.052 (235.5)&0.039 (172.0)\\
							&50\%&0.074 (340.0)&0.112 (525.3)&0.103 (479.1)&0.091 (422.7)&0.063 (286.8)\\
							\hline
							\multirow{2}*{300}&90\%&0.084 (386.1)&0.104 (484.8)&0.086 (394.4)&0.072 (327.6)&0.05 (224.6)\\
							&50\%&0.102 (473.0)&0.145 (693.2)&0.133 (632.9)&0.12 (565.0)&0.082 (377.6)\\
							\bottomrule
						\end{tabular}
						\caption{The detection depths with 90\% and 50\% efficiency calculated by image simulation of WFST with 30, 90 and 300 s exposures.}
						\label{tab:wfst_depth}
			\end{table*}   
			By population synthesis, it is found that most BNS mergers are far from the galaxy center, and the offset distribution depends on gravitational potential, the metallicity of different galaxy types and some parameters of compact binary including formation channel, initial location, kick velocity and delay time \citep{bloom_spatial_1999,belczynski_merger_2002,voss_galactic_2003,belczynski_study_2006,mapelli_cosmic_2018,wang_fast_2020}. By the observation and systematic analysis of the host galaxy, \cite{fong_hubble_2010,fong_decade_2015} and \cite{fong_locations_2013} collected the host galaxies properties of 22 sGRBs and compared their projected offsets (both physical and normalized) with long GRBs, core-collapse SNe, and even Type Ia SNe. They found that most sGRBs are far from the galaxy center, which is consistent with the model where sGRBs originate from the compact binary merger. In addition, the multi-messenger observations of GW170817 further confirmed the relationship between the sGRB and the merger of BNS \citep{goldstein_ordinary_2017}.
			\par  
			We choose to use sGRBs to trace the location of the BNS merger and impose the kilonova into the image according to the projected offset distribution of sGRBs, which means that we assume sGRBs mostly originate from BNS mergers. Besides the 22 sGRBs reported in \cite{fong_hubble_2010,fong_locations_2013}, we collect 5 other sGRBs with reliable offset measurements from 2013 to 2020 and add them into sGRB sample, which can be seen in Table \ref{tab:offset}. The difference between the two samples is shown in Figure \ref{fig:offset}, where the projected physical offsets of the additional five sGRBs are consistent with the distribution of the previous works, and the offset median is 5.1 kpc compared to 4.5 kpc in \cite{fong_locations_2013}. To obtain projected offsets between kilonovae and galaxy centers, we fit this sample of 27 sGRBs with Log-normal distribution using \texttt{statsmodels} \citep{seabold2010statsmodels} based on the maximum likelihood method and the fitting result is:
			\begin{equation} \label{eq:offset}
			\text{PDF}_\text{fit}(r)=\frac{1}{(r-r_0)\sigma\sqrt{2\pi}}\exp \left(-\frac{(\ln(r-r_0)-\mu)^2}{2\sigma^2} \right),
			\end{equation}
	     	where ${\rm PDF}_{\rm fit}$ is a probability density function, $r$ is in kpc and $r_0=0.24\pm0.23$, $\sigma=1.26\pm0.23$ and $\mu=1.44\pm0.27$. Using the PDF (\ref{eq:offset}), the projected offset between the BNS merger's location and the galaxy center can be obtained. Regarding the angular distribution of the merger, we assume a uniform distribution to sample the angle relative to the main axis of the galaxy in our simulation.
	     	\par
	     	\begin{figure}[h]
	     		\centering
	     		{\includegraphics[scale=0.36]{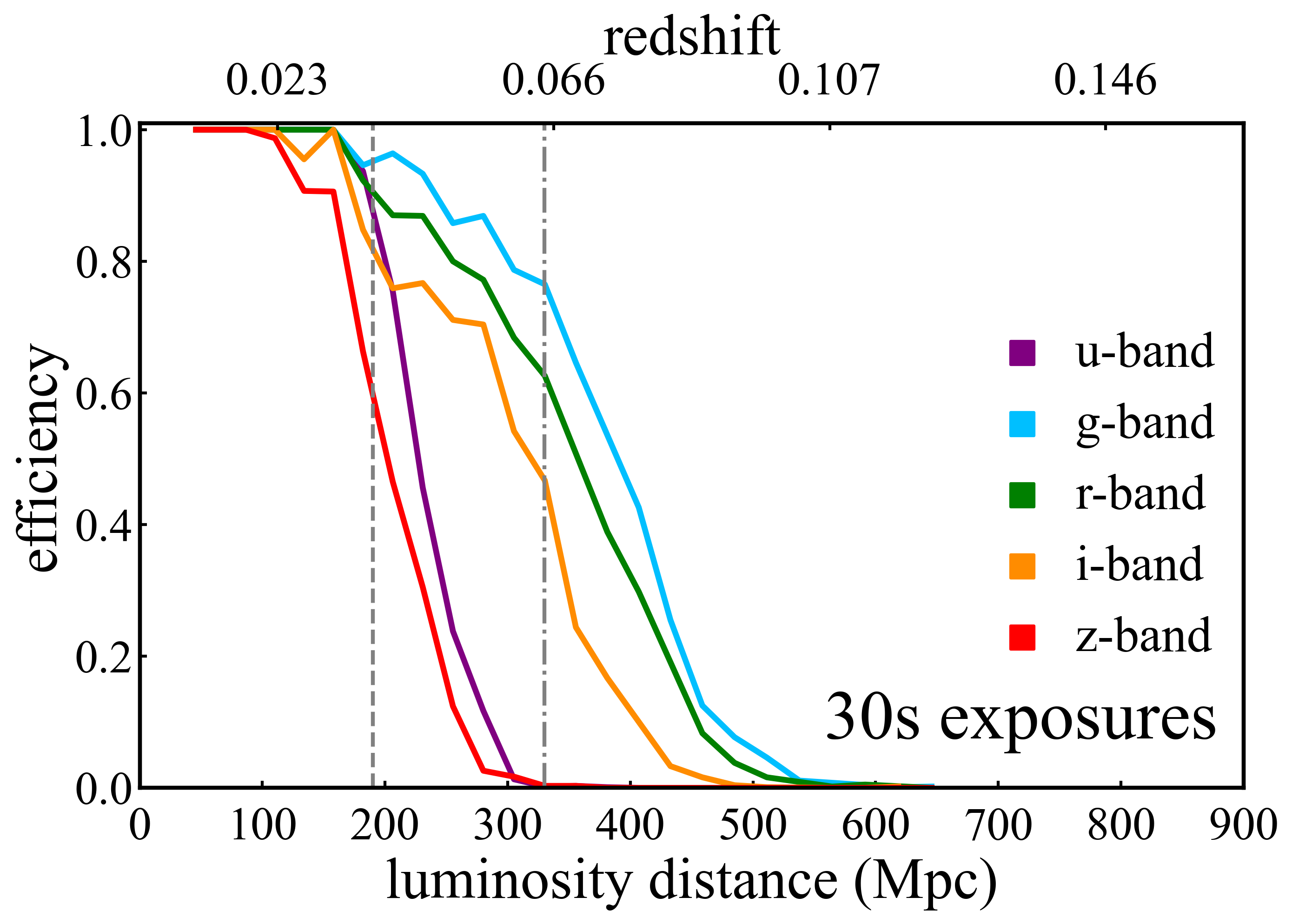} }
	     		{\includegraphics[scale=0.36]{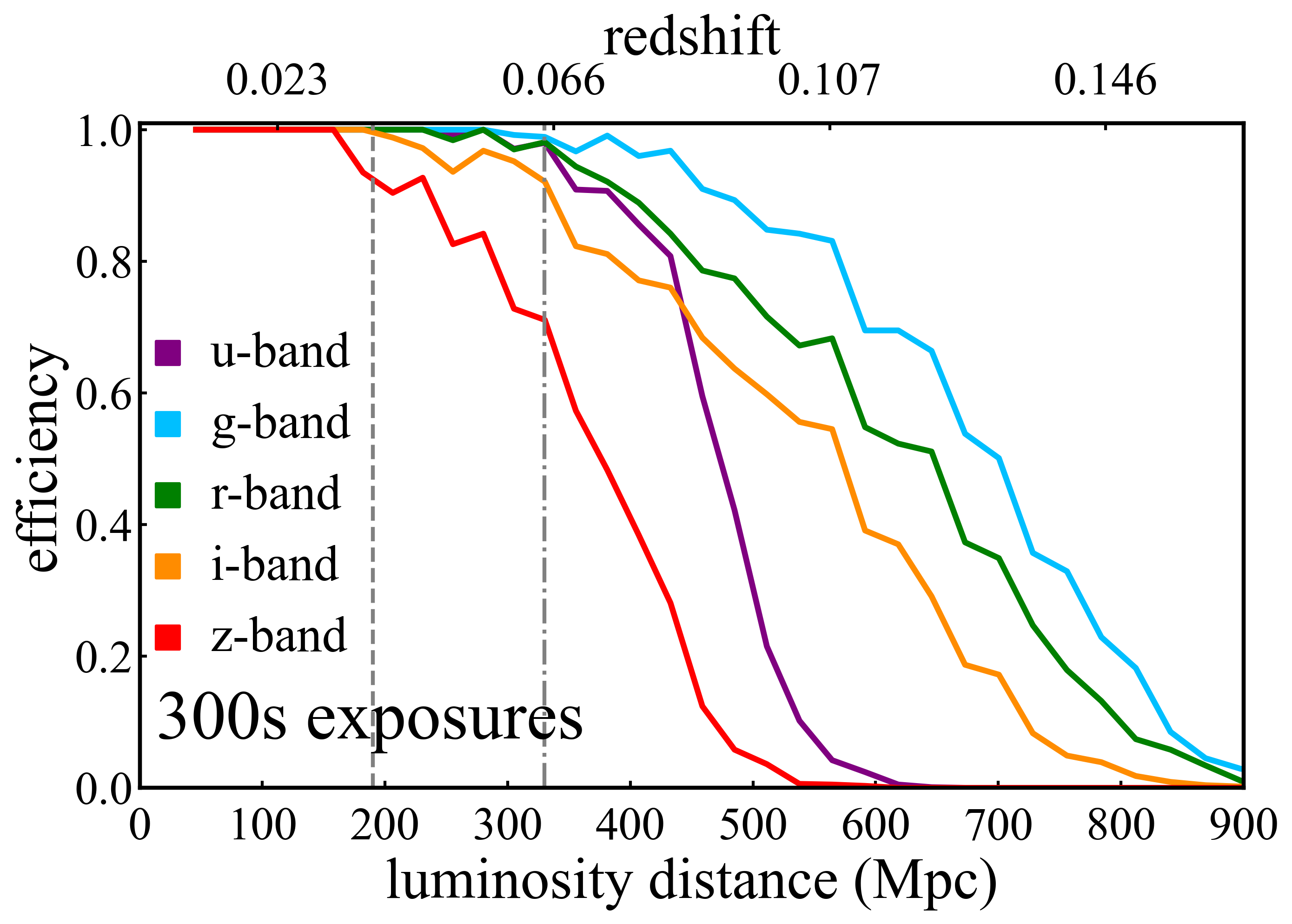} }
	     		\caption{At peak of AT2017gfo-like kilonova, the result of detection efficiency varies with redshift for WFST with 30 (upper) and 300 s (lower) exposures, respectively. The grey dashed and dot-dashed lines represent the detection range for BNS of O4 and O5 \citep{abbott_prospects_2020}.}
	     		\label{fig:wfst_result}
	     	\end{figure}
	        \begin{figure*}[tp]
	     	\centering
	     	\includegraphics[scale=0.45]{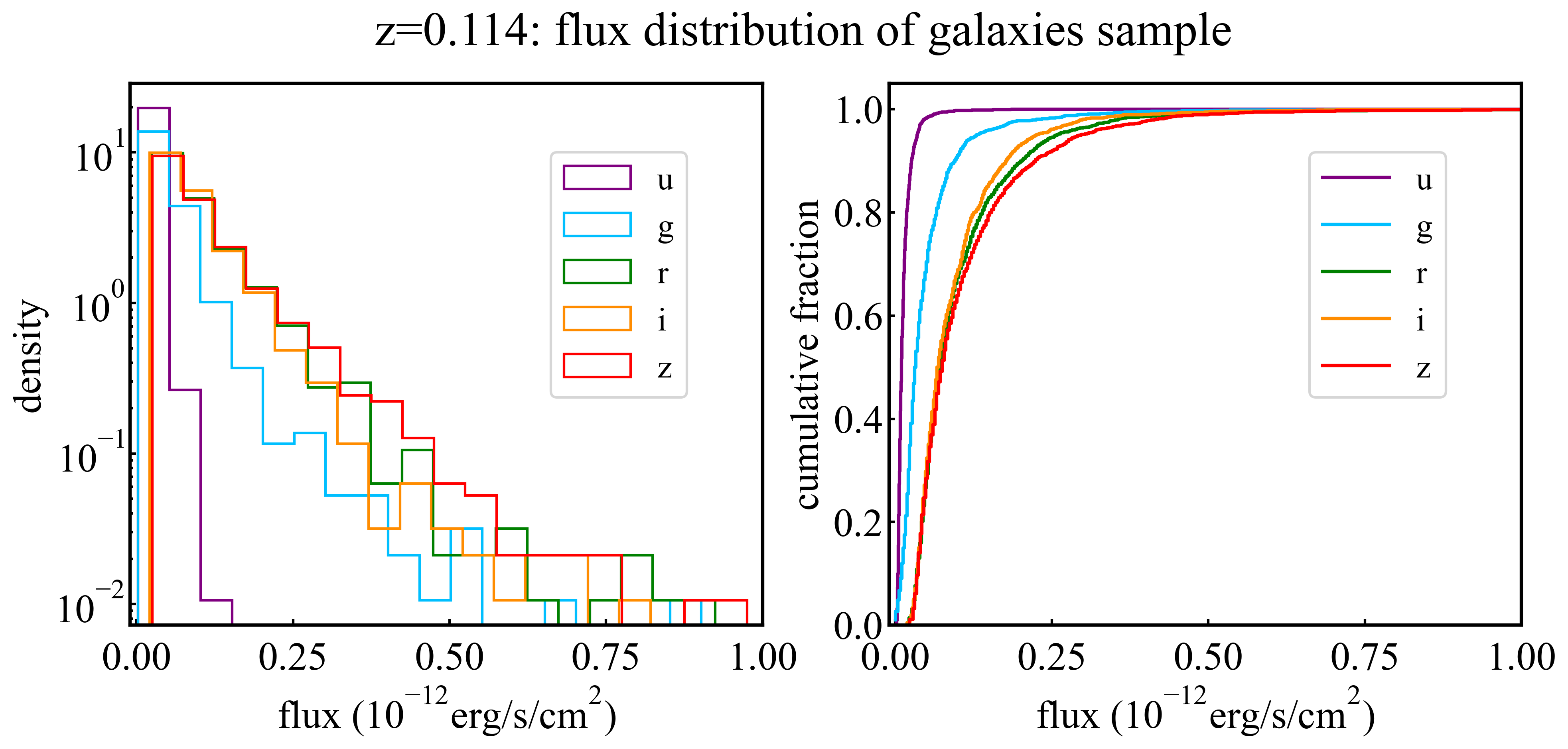}
	     	\caption{The flux distribution of the galaxy sample in CosmoDC2 for each band of WFST at redshift of $z=0.114$.}
	     	\label{fig:distribution}
	     	\end{figure*} 
    	\subsection{Results} \label{result}
	    	We divide the redshift range of $0\sim0.2$ in a grid of 75 bins and generate simulated images for galaxies in each bin. If the number of galaxies in a bin is more than 1000, we will randomly choose 1000 galaxies as the simulation sample to reduce computing time appropriately. An example of simulated images with the host galaxy is presented in Figure \ref{fig:example}. When a BNS merger is close to the galaxy's center, like Panel \ref{subfig:example2}, the background is dominated by host luminosity. Otherwise, in Panel \ref{subfig:example1}, the sky brightness will be vital for detection, and the sky brightness of each band of simulation is shown in Table \ref{tab:teles}. According to the simulation process described in Section \ref{sec_1} and \ref{sec_2}, we generate simulated images for each galaxy in different bands and judge whether kilonovae can be detected or not. By averaging these detection results, the weighted fraction of kilonova that is detectable can be calculated combined with the BNS merger rate derived from Eq (\ref{equation:BNS_rate}). The weighted fraction at different redshifts is called the detection efficiency in our following results. When a AT2017gfo-like kilonova is at the peak luminosity, the results of detection efficiency for WFST with the 30 s and 300 s exposures are shown in Figure \ref{fig:wfst_result}. When a BNS is at small luminosity distances, WFST can easily detect its kilonova emission, as was in the case of GW170817. As the luminosity distance increases, the detection efficiency decreases gradually. The speed of decrease is different for each band, which can reflect the influence of the host galaxy on detection combined with Figure \ref{fig:distribution}. With the subsample at the redshift of $z=0.114$ as an example, Figure \ref{fig:distribution} displays the flux distribution of galaxies of cosmoDC2 in different bands of WFST. According to Figure \ref{fig:distribution}, the longer wavelength of the band, the greater mean and standard deviation of flux distribution. In the u band, there is not much difference in the effect of kilonova detection between different galaxies, because their flux is more concentrated. As a result, the detection efficiency with the u band falls faster than others. In contrast, the efficiency begins to decline earlier and declines more slowly in a longer wavelength band. The detection depths with 90\% and 50\% efficiency are shown in Table \ref{tab:wfst_depth}. Compared with the case of the flat sky brightness in Table \ref{tab:depth}, the detection depth of 90\% efficiency is reduced by $50$ Mpc to $200$ Mpc.
			\par 
	    	\begin{figure}[!h]
	    		\centering
	    		{\includegraphics[scale=0.36]{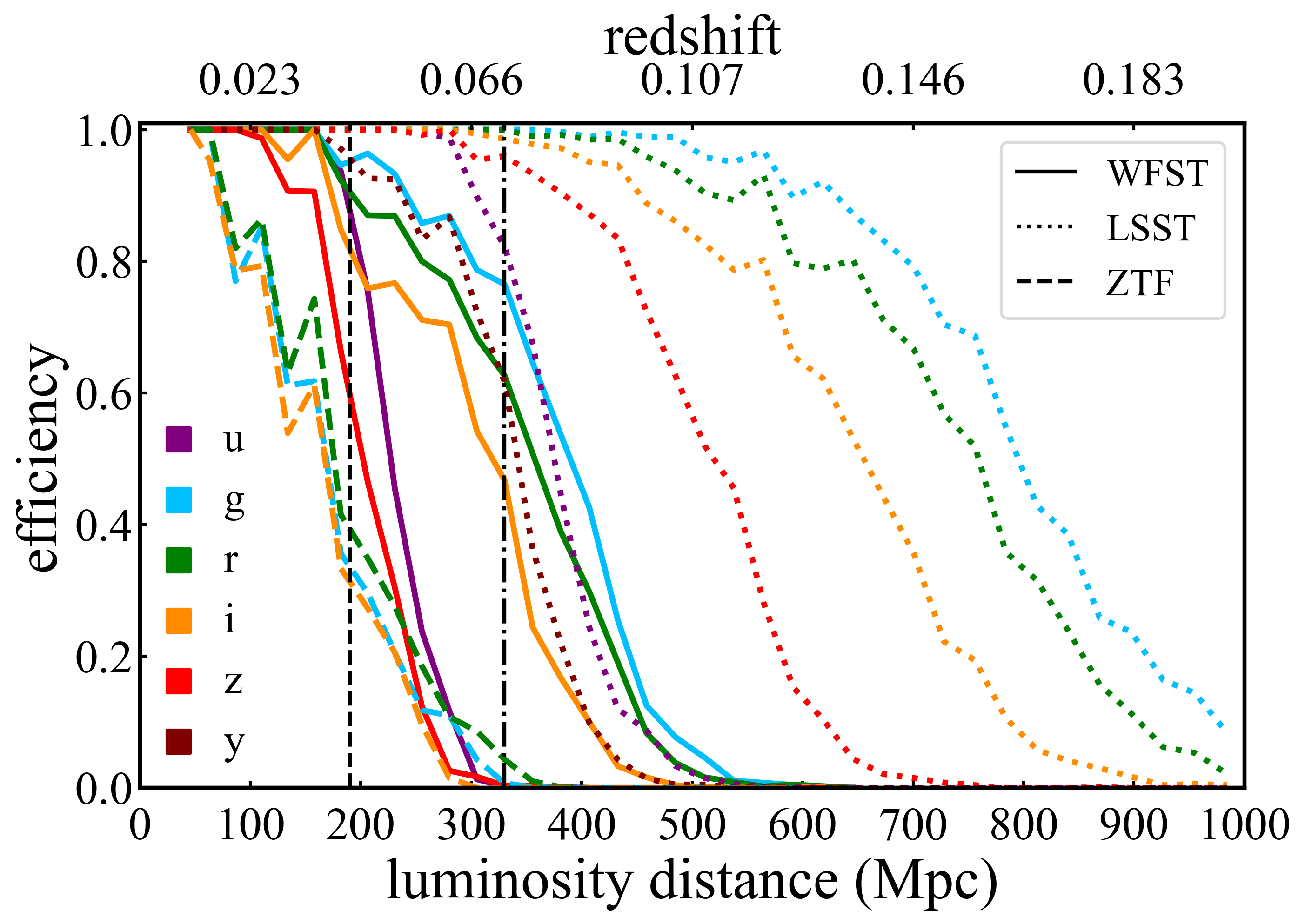} }
	    		\caption{The result of detection efficiency varies with redshift for different telescopes at peak of AT2017gfo- kilonova. For comparing the search ability of kilonova reasonably, exposure time is set as 30, 40 and 200 s for WFST, LSST and ZTF, respectively. The black dashed and dot-dashed lines represent detection range for BNS of O4 and O5 \citep{abbott_prospects_2020}.}
	    		\label{fig:diff_teles}
	    	\end{figure}
			\cite{cowperthwaite_lsst_2019} found that searching kilonova serendipitously with a certain survey is not effective due to its low efficiency if the sensitivity and FoV of the telescope are insufficient. The host galaxy can further affect the detection and measurement of kilonovae and cut down the number of predicted detections. Even under ToO observation, it is necessary to choose the exposure time wisely to balance the coverage of sky area and detection depth, given the rapid evolution of kilonova.
	   		\par
	   		\begin{table*}[htp]
	   			\footnotesize
	   			\begin{tabular}{cccccccc}
	   				\toprule
	   				\multirow{2} *{\bf Telescope}& \multirow{2}*{\bf Efficiency }& \multicolumn{6}{c}{ $\bm{z_{\rm max}\;(D_{\rm L},_{\rm max}}$ \bf Mpc)}\\
	   				\cline{3-8}&&\bf u&\bf g&\bf r&\bf i&\bf z&\bf y\\
	   				\midrule
	   				\multirow{2}*{WFST}&90\%&0.039 (175.3)&0.055 (246.7)&0.046 (205.8)&0.038 (168.4)&0.036 (158.5)&\multirow{2}*{...}\\
	   				&50\%&0.051 (230.4)&0.084 (386.7)&0.078 (357.3)&0.069 (316.0)&0.045 (202.0)\\
	   				\hline
	   				\multirow{2}*{LSST}&90\%&0.067 (304.6)&0.131 (623.7)&0.113 (532.2)&0.098 (454.3)&0.084 (387.7)&0.054 (242.1)\\
	   				&50\%&0.082 (375.1)&0.165 (799.3)&0.156 (751.5)&0.137 (653.7)&0.11 (517.1)&0.074 (340.4)\\		
	   				\hline
	   				\multirow{2}*{ZTF}&90\%&\multirow{2}*{...}&0.018 (78.7)&0.019 (84.5)&0.016 (71.2)&\multirow{2}*{...}&\multirow{2}*{...}\\
	   				&50\%&&0.037(165.3)&0.04(180.5)&0.036(159.0)&&\\
	   				\bottomrule
	   			\end{tabular}
	   			\caption{The detection depths with 90\% and 50\% efficiency calculated by image simulation of WFST, LSST and ZTF. Exposure time is set as 30, 40 and 200 s for WFST, LSST and ZTF, respectively.}
	   			\label{tab:depth_diff_tele}
	   		\end{table*}
	   		\begin{figure*}[ht]
	   			\centering
	   			{\includegraphics[scale=0.4]{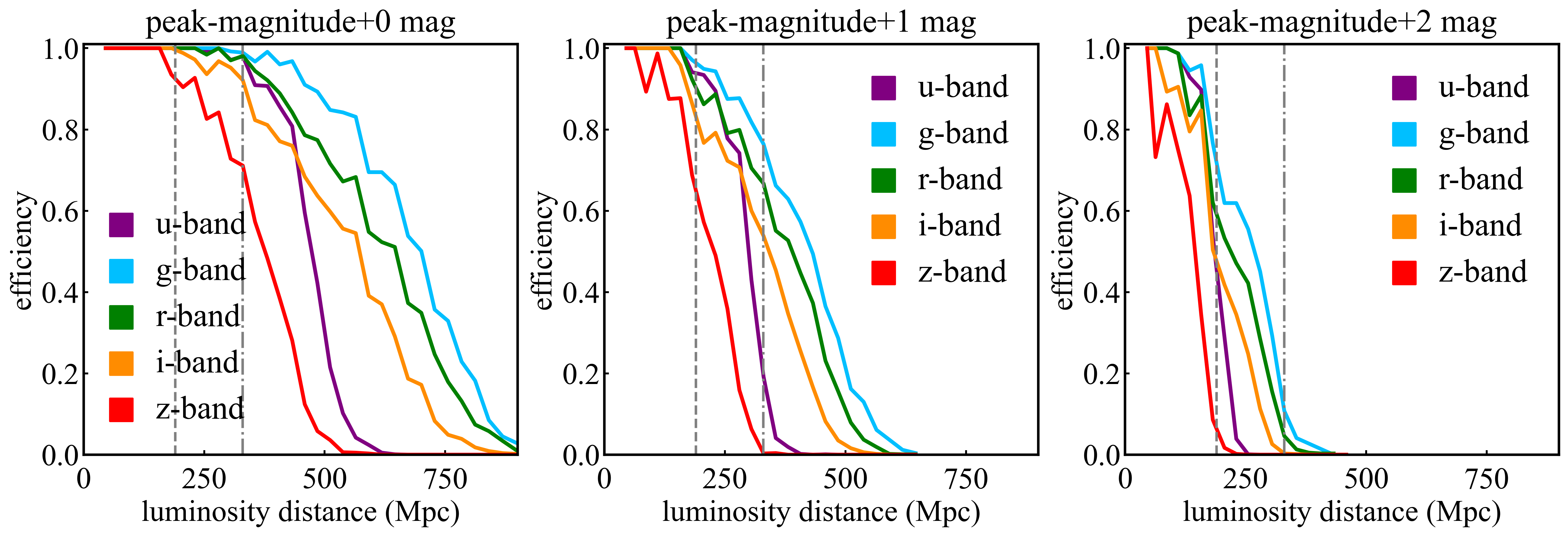} }
	   			\caption{At different times of evolution of AT2017gfo-like kilonova, the detection efficiency for WFST with 300 s exposures. The grey dashed and dot-dashed lines represent detection range for BNS of O4 and O5 \citep{abbott_prospects_2020}.}
	   			\label{fig:diff_mag}
	   		\end{figure*}
	   		We also compute and simulate the situations of the other wide-field surveys which are dedicated to fast-transient discovery: LSST \citep{lsst_science_collaboration_science-driven_2017,ivezic_lsst_2019} and ZTF \citep{bellm_zwicky_2019,graham_zwicky_2019}. The sky brightness of the sites and other parameters of telescopes used in simulation are shown in Table \ref{tab:teles}. The contrast of detection efficiency of these projects with WFST is presented in Figure \ref{fig:diff_teles} and their detection depths of 90\% and 50\% efficiency are also shown in Table \ref{tab:depth_diff_tele}. Because of the different FoV of these telescopes, in order to reasonably reflect their detection abilities for AT2017gfo-like kilonovae, their exposure times need to be different according to their FoV. Assuming the same total coverage of sky area and searching time, the exposure time used is inversely proportional to the FoV size under full coverage. Hence, we set the exposure times as 30, 40 and 200 s for WFST, LSST and ZTF, respectively. From Table \ref{tab:teles} and Table \ref{tab:depth_diff_tele}, although the FoV of ZTF is much larger than WFST, for distant BNSs, WFST can perform better than ZTF and increase the search depth to $200\sim300$ Mpc with 90\% detection efficiency. By simulation in the Southern Hemisphere, LSST is expected to be more powerful and can reach $\sim600$ Mpc $(z=0.13)$ for searching kilonovae. Our result is consistent with the detectability study presented by \cite{scolnic_how_2017}, which predicted a detectable redshift range of $z = 0.02\sim0.25$ for LSST based on an AT2017gfo-like model. Given the geographical locations of WFST and LSST, WFST can complement LSST for optical follow-up in the northern hemisphere well to cover GW events of all-sky.
	   		\par
			To guide the selection of exposure time in searching kilonovae, we also consider the rapid decay of kilonova luminosity. In the case of one or two magnitudes fainter than peak luminosity, which roughly corresponds to the $\sim$2nd and $\sim$3rd day after the merger \citep{kasliwal_kilonova_2020}, their results are shown in Figure \ref{fig:diff_mag}. When luminosity decays one magnitude from the peak, the detection depth with the same efficiency decreases a lot, which indicates that it is essential to make the best of hours around the peak and increase the exposure time in the following days after the peak. 
    \section{Target of Opportunity optimization with WFST} \label{sec_3}
		In this section, based on the simulation, we investigate the ToO observation optimization for WFST and discuss the follow-up ability and detection prospects with WFST for upcoming O4.
		\par
    	\subsection{Optimize the exposure time}
	    	After a GW alert triggers, brief information about the GW event is expected to be released by the Gamma-ray Coordinates Network platform. Then, the follow-up observation will begin based on the BAYESTAR sky map or the LALInference sky map which include more detailed information about the location and distance. Due to the different methods used, the difference between these sky maps is the computing time and estimation accuracy, and the BAYESTAR sky maps are always released earlier but have worse estimations. By analyzing the \texttt{fits} file of sky map, the probability distribution of location and luminosity distance can be obtained and used for ToO observation design. Combining the luminosity distance from GW alert and the ability of the telescope, we can choose an appropriate exposure time so that WFST can strike a balance between depth and coverage.
	    	\par
	     	About luminosity distance in the \texttt{fits} file of sky map, there is detailed introduction in \cite{singer_going_2016,singer_supplement_2016}. The sample of luminosity distance follows the distribution: 
	     	\begin{equation}
	     		\begin{aligned}
	     		p(r|\boldsymbol{n}) & =\frac{N(\boldsymbol{n})}{\sqrt{2 \pi}\sigma ( \boldsymbol{n})}\exp{\bigg[-\frac{(r-\mu(\boldsymbol{n}))^2}{2\sigma (\boldsymbol{n})^2}\bigg]}r^2 \\ 
	     		& \text{for } r \ge 0,
	     		\end{aligned}
	     	\end{equation}
	     	where $\boldsymbol{n}$ is a direction vector, which means the direction to a certain pixel of the sky based on HEALPix (Hierarchical Equal Area isoLatitude Pixelisation) algorithm, and parameters $N$, $\mu$ and $\sigma$ are stored in the \texttt{fits} file. We choose the distance $r_{\text{thred}}$ with 90\% quantile of the distribution as the merger's location. By taking into account the result of detection efficiency in Section \ref{sec_3}, the exposure time is selected from an optional time list set as [30, 60, 90, 120, 180, 240, 300 s], and we require the efficiency is no less than 90\% at $r_{\rm thred}$. If the maximum exposure in the list is insufficient to meet the threshold efficiency, the exposure time is set at 300 s. For the detection efficiency, there are three related parameters: exposure time, kilonova brightness and bandpass. Given the occurrence time of the merger, we use interpolation to generate the detection efficiency curve under other exposure times and the average magnitude of each observable time which can be calculated using \texttt{Astropy}. Thus, we can reasonably set different exposure times for the telescope according to various events and observation nights.
	     	\par
    	\subsection{Follow-up ability and prospect} 
			Determining how much time is needed to cover the probable sky region in a follow-up observation can reflect the feasibility to search for kilonovae and guide us to allocate observation time for various GW events. However, in practice the total time of a follow-up observation is usually affected by too many factors, e.g., occurrence time, localization accuracy, weather and Moon phase. To reflect the ability of WFST to follow-up search in O4, we investigate the average total time in ToO observation under various localization areas and distances.
	    	\par
			The probable localization is assumed to be ideal enough, which means the observation is not influenced much by the Moon, the foreground of the Milky Way, and the weather. Combining the localization area of the event and the telescope's FoV, the number of images taken can be roughly estimated. By sampling the trigger time of GW events randomly in a day, based on the selection rule above, the exposure time for each observable time can be determined. We can subsequently match the exposure time to each pointing of the telescope and calculate the total time for covering the target sky. According to the prospects of the localization of GW events in \cite{abbott_prospects_2020}, the median localization area of BNS from LHVK measurements in O4 is $33^{+5}_{-5} \, \text{deg}^2$. Since the localization accuracy is sensitive to the SNR of signal and the number of detectors, the localization area with 90\% confidence of most events might be hundreds to even thousands of square degrees in O4 \citep{petrov_data-driven_2022}. Consequently, we set the calculation range of luminosity distance and localization area as $0 \sim 500 \, \text{Mpc}$ and $0 \sim 1250 \, \text{deg}^2$, respectively. 
	    	\par
\begin{figure*}[t]
			    	\centering
			    	\subfigure[]{	
			  			\label{subfig:1}
			    		\includegraphics[scale=0.37]{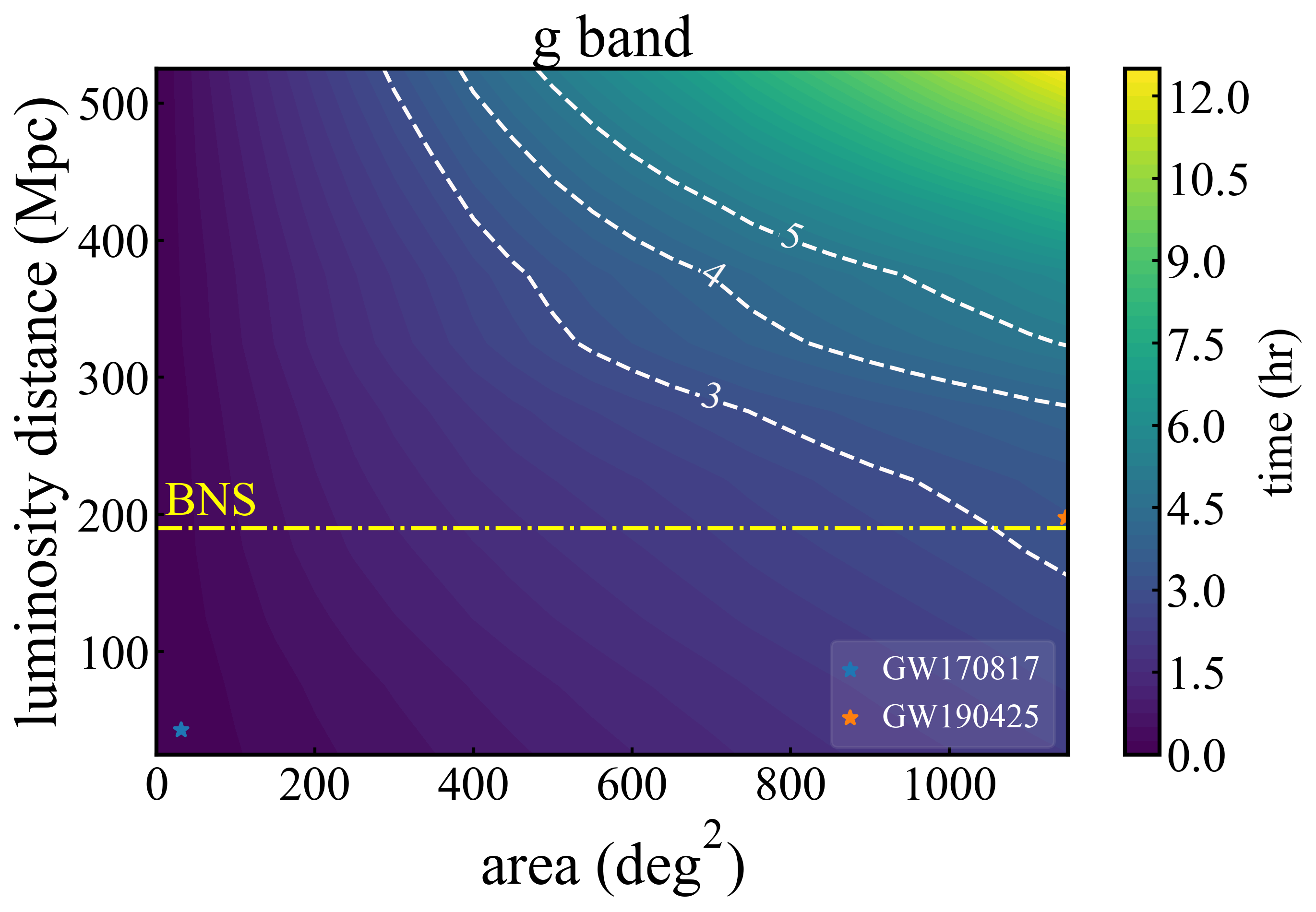}
			    	}
			   		\subfigure[]{	
						\label{subfig:2}
			   			\includegraphics[scale=0.37]{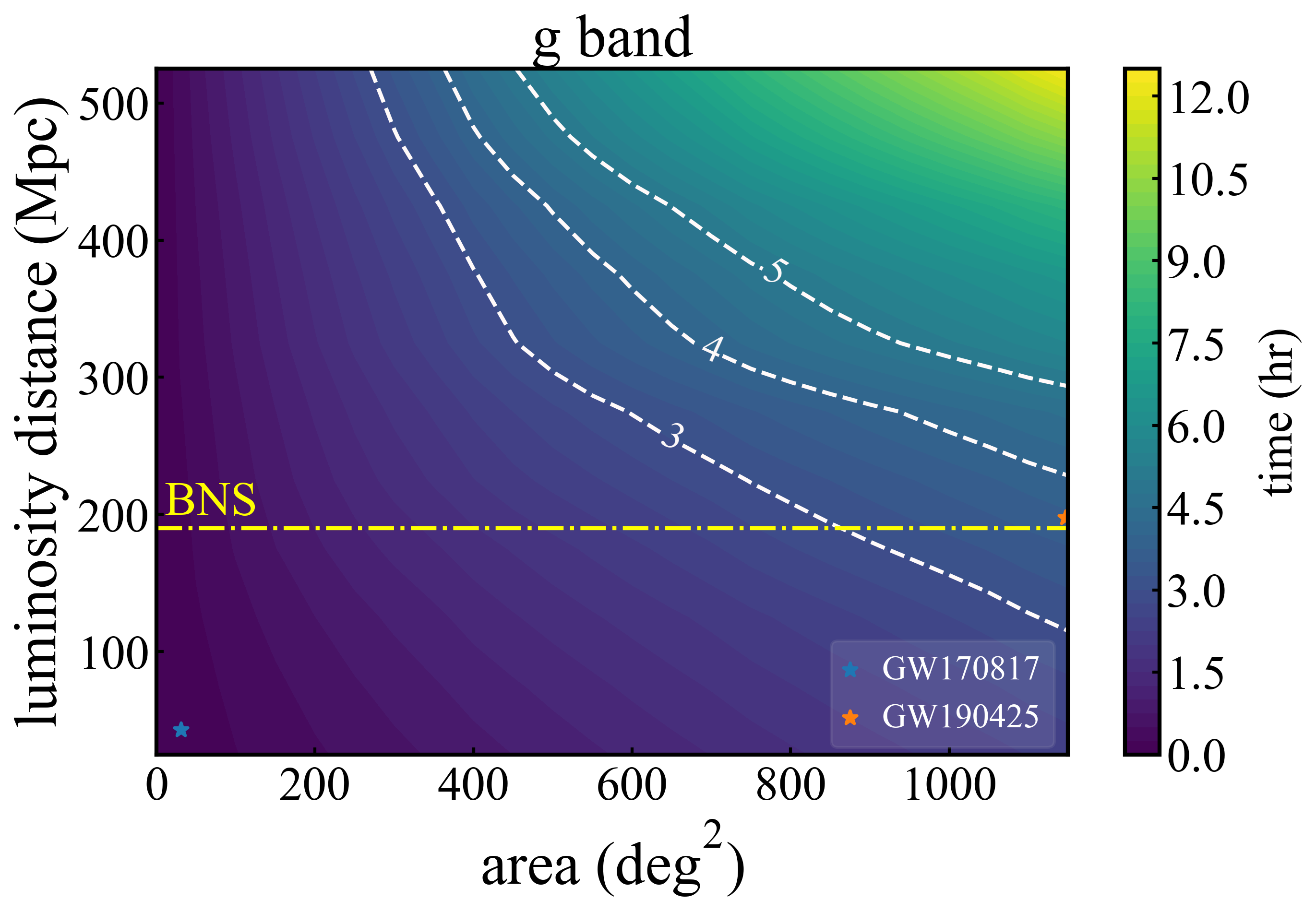}
			   		}\\
					\subfigure[]{
						\label{subfig:3}
						\includegraphics[scale=0.37]{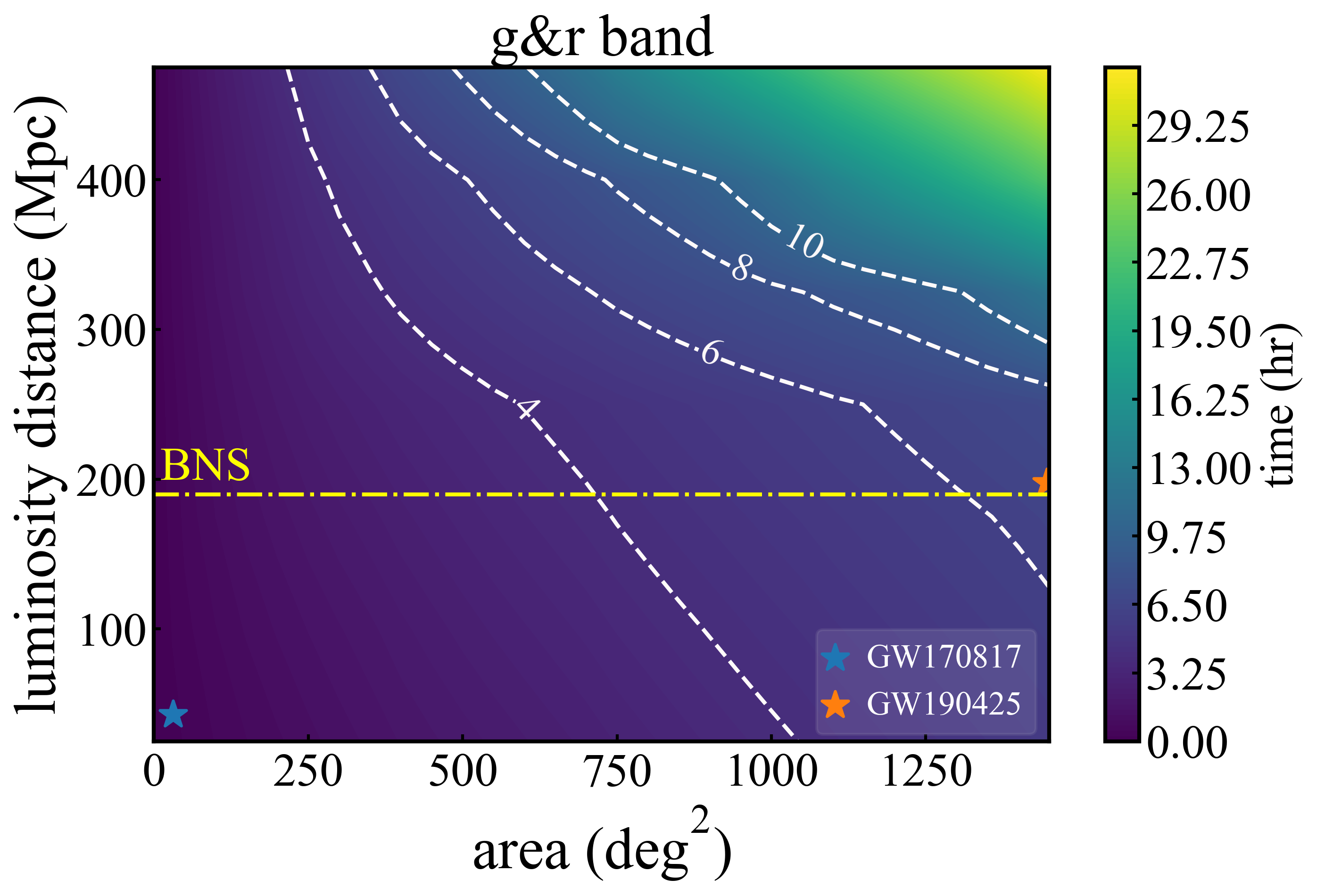}
					}
					\subfigure[]{
						\label{subfig:4}
						\includegraphics[scale=0.37]{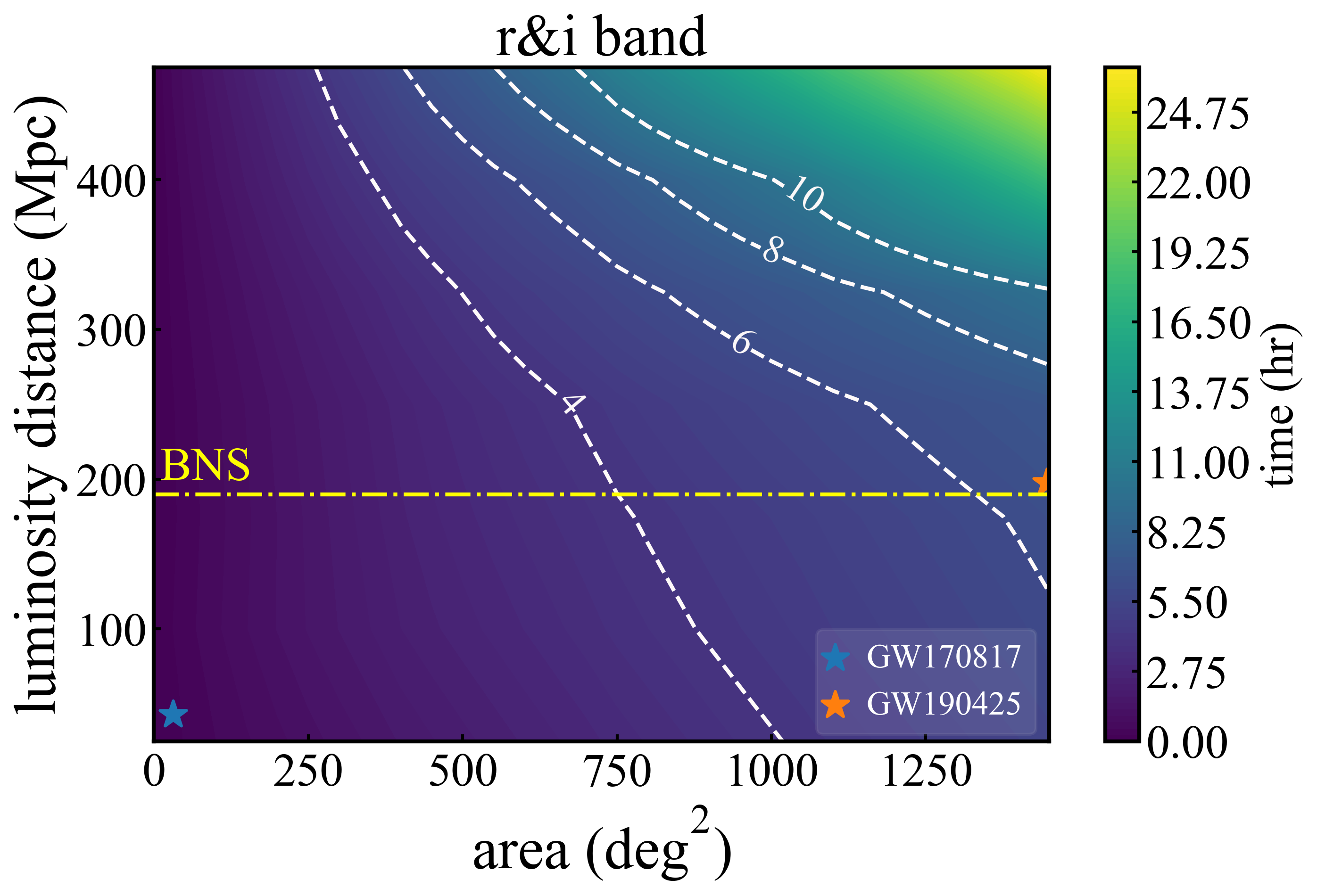}
					}
			    	\caption{The results of total time in ToO observations with different localization and luminosity distance. The dates of \subref{subfig:1}, \subref{subfig:3} and \subref{subfig:4} are set to 2023.06.21 and \subref{subfig:2} is 2023.12.23. The dot-dashed line annotates the limit distance for detecting the typical BNS merger in O4 \citep{abbott_prospects_2020}. White dashed lines represent contour lines of different times. GW170817 and another possible BNS event GW190425 detected in O3 are also drawn in each panel, represented by star markers. Their distances are taken as $\mu+\sigma$, where $\mu$ and $\sigma$ are the median and standard deviation, and the localization area is the part with 90\% confidence where ${\rm decl.}>-30^{\circ}$. The event that is out of the calculation range is drawn on the axes.}
			    	\label{fig:ToO}
    		\end{figure*}

	    	The results of total time with only one band and two bands are shown in Figure \ref{fig:ToO}. To investigate the influence of observable time per night of different seasons, we set two calculation dates to 2023.06.21 or 2023.12.23 to represent the case of summer or winter. Comparing Panel \ref{subfig:1} and \ref{subfig:2}, there is not much difference in total time between summer and winter at the Lenghu site, but the target sky can be covered earlier in winter due to the more time per night. To further filter and confirm the kilonova candidate, using at least two bands is necessary for getting color evolution in actual observation. Therefore, the results of two-band coverage in summer are also calculated and shown in Panel \ref{subfig:3} and \ref{subfig:4}. GW170817 and another possible BNS event GW190425 reported during O3 are also marked in each panel. GW190425 is placed on the axes due to its bad localization. For the band choice, the combination between g , r  and i bands are preferred for their relatively higher efficiency to search for kilonovae. In addition, a bright Moon phase can affect a lot on shorter wavelength in optics, especially for the u and g bands of WFST. Referring to the GW ToO strategy of LSST \citep{andreoni_target_2021}, we choose the g and r bands in dark times of the Moon phase and a longer wavelength band combination r and i bands in bright times, which correspond to Panel \ref{subfig:3} and \ref{subfig:4} respectively. The total time spent using r and i bands is slightly longer than g and r bands, which is predictable according to our optimization method of exposure time. Additionally, the limiting magnitude can also be reduced depending on the Moon phase as well as the angular distance between the pointing of the telescope and the Moon. Therefore, in practice appropriately avoiding the sky close to the Moon is necessary to ensure overall efficiency. Assuming that the observation time per night is $\sim4 \, \text{hr}$, and it is more promising to find the kilonova in the first two nights, we find WFST can search for kilonovae well if the localization area is no more than ${10}^3 \, {\deg}^2$, given the detection limit of GW detectors in O4. If the localization accuracy is poor $(>{10}^3 \, {\deg}^2)$, as reported for some events in O3, it is necessary to give up part of the area and focus on areas with relatively high probability. Based on this result, if the localization area of most BNS events detected in O4 is hundreds of square degrees, WFST can detect most kilonova counterparts of GW events. Assuming WFST can observe the sky where $\text{decl.}>-30^{\circ}$, amongst kilonovae that are more than 15 degrees from the disk of the Milky Way, only 57\% events can be searched by WFST. Combined with the expected BNS detections $10^{+53}_{-10}$ during O4 in \cite{abbott_prospects_2020} and weather efficiency of 76\% of the site \citep{deng_lenghu_2021}, roughly considering the influence of the Moon and the Sun, WFST is expected to find $\sim 30\%$ ($3^{+16}_{-3}$) kilonovae from BNS mergers during O4.
    		\par
    \section{Discussions and conclusions} \label{sec_4}
        \subsection{The influence of different kilonova models} \label{other kilnovae}
The detection efficiency of the telescope depends on the kilonova models. However, the physical model of kilonova is still unclear, even for a given BNS parameter, the viewing angle and the EoS of neutron stars. Since only one event, i.e. GW170817, is confirmed until now, various models are proposed in the literature, and most of them can explain the data fairly well (see for instance \cite{review3, review1,review2}). In observations, there are six sGRBs with potential kilonova emission detected, with different levels of observational evidence. These sGRB kilonovae support the existence of a broad range of kilonova luminosities ($\sim 0.3-10$ times the luminosity of AT2017gfo depending on the epoch and frequency of observations) \citep{ascenzi_luminosity_2019}, which might hint the diversity of kilonova models \citep{review2}. In the previous discussion, we ignored these model uncertainties, and simply assumed that all the events are AT2017gfo-like. In order to investigate the effects of different kilonova models on detection efficiency, we simulate a sample of BNS models and consider the alternative kilonova model to calculate the luminosity of the events.

In fact, the luminosity of kilonovae can be different depending on the properties of BNS systems. Based on the results of different kilonova luminosities in Figure \ref{fig:diff_mag}, combined with a kilonova sample generated from various BNS systems, the detection efficiency can be estimated simply for kilonovae deviating from the AT2017gfo-like kilonovae. To construct a kilonova sample, we refer to the method in \cite{wanghuiyu_2022} and introduce the method as follows.
            \par
			To obtain the corresponding kilonova of each BNS system in the sample, we use the kilonova model (\texttt{bns} model) in \cite{nicholl_tight_2021}. Similar to the previous model in this paper, the \texttt{bns} model is also composed of three-component ejecta matter which includes blue, purple and red components. These components correspond to different generation mechanisms, because of which the opacity and atomic mass number of diverse components are different. In contrast to the previous model, the geometry of ejecta is axisymmetric in the \texttt{bns} model, where the blue and purple components mainly distribute around the pole direction and the red component concentrates around the equatorial plane, so the influence of the observer viewing angle on light curves can be considered. Using the fitting results of relativistic numerical calculation in \cite{dietrich_modeling_2017}, the authors connected the parameters of each ejecta matter component with the mass ratio and chirp mass of a BNS system. The mass and velocity of the ejecta matter can be calculated from the masses of BNS. In addition, the other two effects that can contribute additional luminosity to kilonovae are also considered: (1) an enhancement in the blue ejecta due to magnetically-driven winds and (2) the shock-heating of the ejecta by a GRB jet, where (1) can be possible only if the remnant avoids prompt collapse \citep{Metzger_2018}, and (2) can contribute to blue luminosity at the early evolution of kilonovae \citep{arcavi_first_2018}.
			\par
			The free parameters of \texttt{bns} model fitting AT2017gfo are: mass ratio $(q)$, chirp mass $(\mathcal{M})$, symmetric tidal deformability $(\Lambda_{s})$, redshift $(z)$, observer viewing angle $(\theta)$, enhancement factor of blue ejecta by surface winds $(\alpha)$, fraction of disk ejected $(\epsilon_{\rm disk})$, maximum stable NS mass $(M_{\rm TOV})$ and opening angle of shocked cocoon $(\theta_{\rm c})$. The mass ratio and the chirp mass can be derived from the NS mass by equations:
			\begin{equation}
				q=\frac{M_1}{M_2} \le 1,
			\end{equation}
			\begin{equation}
	\mathcal{M}=\frac{{(M_1 M_2)}^{{3}/{5}}}{{(M_1+M_2)}^{{1}/{5}}}.
			\end{equation}
			The observer viewing angle of each BNS is sampled uniformly from $\pi/2$ to $-\pi/2$. To directly compare these lightcurves with AT2017gfo, the redshift is set as $z=0.0098$, which is the same as AT2017gfo \citep{soares-santos_electromagnetic_2017}. The symmetric tidal deformability can be calculated from $q$ and $\mathcal{M}$. The symmetric tidal deformability $\Lambda_s$ and the antisymmetric tidal deformability $\Lambda_s$ is defined as $\Lambda_s \equiv \Lambda_1 + \Lambda_2$ and $\Lambda_a \equiv \Lambda_1 - \Lambda_2$, where $\Lambda_1$ and $\Lambda_2$ are external tidal fields of BNS. The relationship between $\Lambda_s$ and $\Lambda_a$ is given in \cite{Yagi_2016}:
			\begin{equation} \label{eq:Yagi}
				\Lambda_{a} = F_{\bar{n}}(q) \Lambda_{s} \frac{a+\sum^{3}_{i=1} \sum^{2}_{j=1} b_{ij} q^j \Lambda_{s}^{-i/5}}{a+\sum^{3}_{i=1} \sum^{2}_{j=1} c_{ij} q^j \Lambda_{s}^{-i/5}},
			\end{equation}
			\begin{equation} \label{eq:Yagi1}
				F_{n}(q) \equiv \frac{1-q^{10/(3-n)}}{1+q^{10/(3-n)}},
			\end{equation}
			where $a=0.0755$ and $\bar{n}=0.743$, which means the average factor of various EoS. $b_{ij}$ and $c_{ij}$ are listed in Table \ref{tab:matrix}. A specific combination of $\Lambda_{1}$ and $\Lambda_{2}$ can be constrained by GW information \citep{PhysRevD.89.103012,PhysRevLett.112.101101}:
			\begin{equation} \label{eq:lambda_bar}
				\widetilde{\Lambda} = \frac{16}{13} \frac{(12q+1)\Lambda_{1}+(12+q)q^4 \Lambda_{2}}{(1+q)^5},
			\end{equation}
			called the binary or effective tidal deformability, which can also be calculated according to the empirical relationship with the radius of a 1.4$M_\odot$ NS $(R_{1.4})$:
			\begin{equation} \label{eq:lambda_emp}
				\widetilde{\Lambda} \simeq 800 \left(\frac{R_{1.4}}{11.2} \frac{M_{\odot}}{\mathcal{M}} \right)^6.
			\end{equation}
			Based on a certain $R_{1.4}$ and $M_{\rm TOV}$, $\Lambda_s$ can be calculated combined with the Eqs (\ref{eq:Yagi}), (\ref{eq:Yagi1}), (\ref{eq:lambda_bar}) and (\ref{eq:lambda_emp}). The undetermined parameters are set referring to the best fitting result of AT2017gfo in \cite{nicholl_tight_2021} as: $\epsilon_{\rm disk}=0.12$, $\alpha=0.63$, $M_{\rm TOV}=2.17 M_{\odot}$, $R_{1.4}=11.06 \,$km and $\cos \theta_{\rm c}=0.91$.
            \par
            We refer to the results in \cite{farrow_mass_2019} to construct a BNS sample. In the standard isolated binary formation channel, a recycled neutron star (NS) is born first and spins up to $\sim$ 10--100 ms by an accretion/recycling process. Its companion a nonrecycled NS quickly spins down to $\mathcal{O}(1)$ second period after birth \citep{tauris_formation_2017}. By fitting 17 Galactic BNS systems, in the best fitting case, \cite{farrow_mass_2019} found that the nonrecycled NS mass is distributed uniformly, and the recycled NS mass distributes according to a two-Gaussian distribution: 
			\begin{equation}
				\begin{aligned}
				&\pi (m|{\mu_1,\sigma_1,\mu_2,\sigma_2,\alpha})=\frac{\alpha}{\sigma_1 \sqrt{2\pi}} \times \\ 
				&\exp \left[-\left(\frac{m-\mu_1}{\sqrt{2}\sigma_1}\right)^2 \right]+\frac{1-\alpha}{\sigma_2 \sqrt{2\pi}}\exp \left[-\left(\frac{m-\mu_2}{\sqrt{2}\sigma_2}\right)^2 \right]\\
				\end{aligned}
			\end{equation}
			where the fitting parameters are: $\mu_1=1.34M_{\odot}$, $\mu_2=1.47M_{\odot}$, $\sigma_1=0.02M_{\odot}$, $\sigma_2=0.15M_{\odot}$ and $\alpha=0.68M_{\odot}$. For the nonrecycled neutron star, the range of uniform distribution is 1.16--1.42$M_{\odot}$. According to this fitting result, the NS masses of the BNS system can be sampled.
			\par
\begin{figure}[ht]
				\centering
				{\includegraphics[scale=0.5]{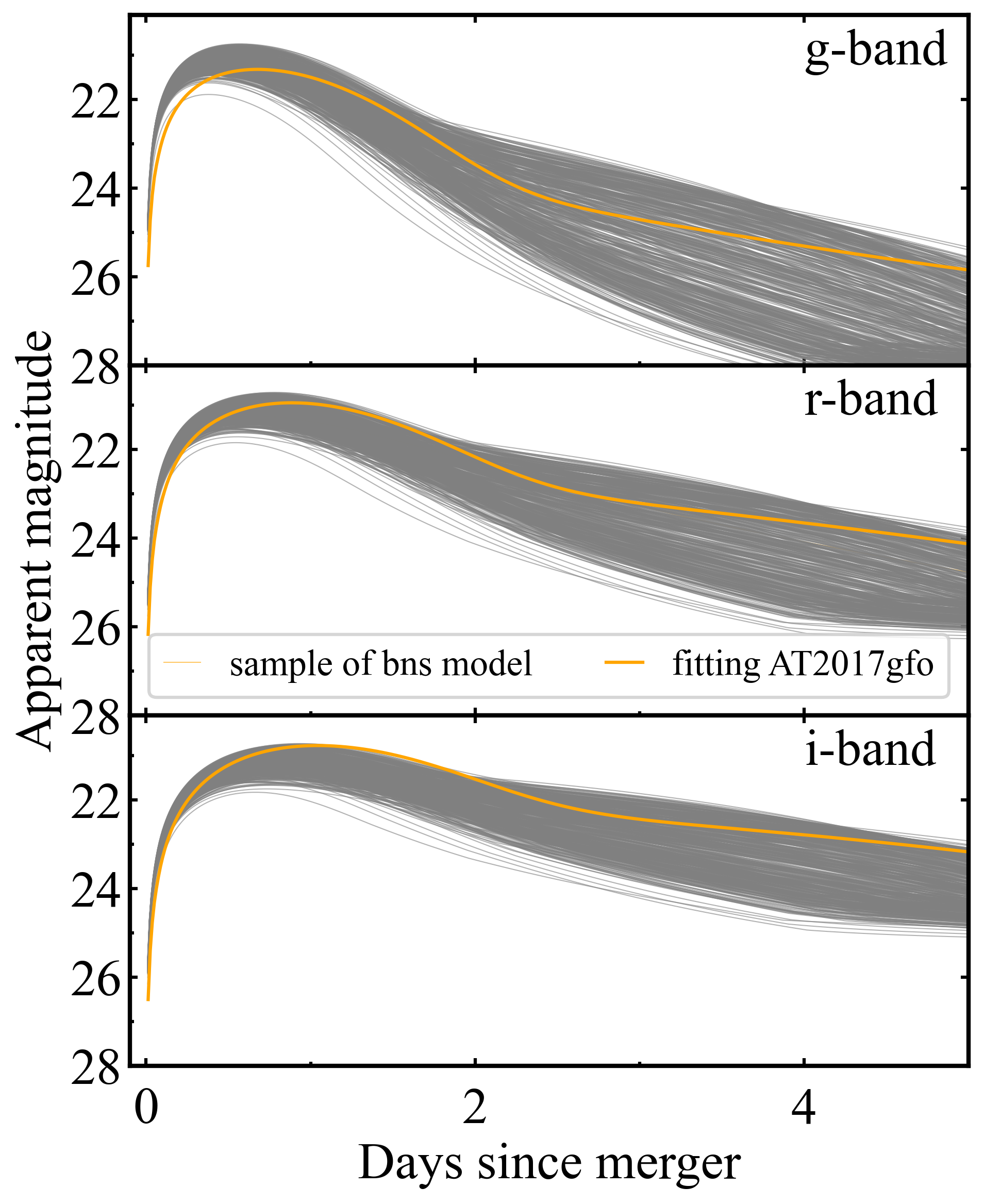}}
				\caption{The lightcurves calculated from bns model (grey) and fitting AT2017gfo lightcurves (orange) in g, r and i bands of WFST. Their redshifts are set as the same with AT2017gfo.}
				\label{fig:bns_sample}
			\end{figure}

			\begin{table}
				\centering
				\begin{tabular}{*{8}{c@{\hspace{0.2cm}}}}
					\toprule
					$b_{ij}$ & $i=1$ & $i=2$ & $i=3$ & $c_{ij}$ & $i=1$ & $i=2$ & $i=3$\\
					\midrule
					$j=1$ & -2.235 & 10.45 & -15.75& $j=1$ & -2.048 & 7.941 & -7.36\\
					$j=2$ & 0.847 & -3.25 & 13.61& $j=2$ & 0.598 & 0566 & -1.32\\
					\bottomrule
				\end{tabular}
				\caption{The value of parameters in Eq \ref{eq:Yagi}.}
				\label{tab:matrix}
			\end{table}
			Combined with the BNS sample and \texttt{bns} model, in g, r, i bands of WFST, the lightcurves of the sample of 500 BNS systems are calculated and shown in Figure \ref{fig:bns_sample} as a cluster of grey lines. As contrast, the light curves of fitting AT2017gfo kilonova are also drawn as orange lines. From Figure \ref{fig:bns_sample}, the evolution of AT2017gfo can be included well in our sample. Compared with the AT2017gfo, a kilonova in the sample has a lower luminosity than AT2017gfo in most cases, which is broadly consistent with the result in \cite{chase_kilonova_2022}, where the authors generated a kilonova sample based on another model (SuperNu) to investigate the kilonova detectability. In Figure 3 of \cite{chase_kilonova_2022}, in the r band of LSST, the detection depth of AT2017gfo is farther away than most of the sample, which means that AT2017gfo is more luminous in the sample. In the first two days, which is the most promising time to detect the kilonova, the magnitude difference between the kilonovae sample and AT2017gfo is $-0.5$ to 1 mag. For an extreme case that a kilonova is 1 mag fainter than AT2017gfo at peak luminosity, the detection depths (observed by WFST with 90\% efficiency and 300 s exposure time ) decrease to $\sim 200$ Mpc according to Figure \ref{fig:diff_mag}. Therefore, considering the luminosity function of kilonovae, the average predicted detection range can further decrease compared to the AT2017gfo-like kilonovae.
		\subsection{Conclusion}
	    This work explores the impact of image subtraction and host galaxy interference on WFST's ability to detect kilonovae and uses these results to optimize the exposure selection for future ToO observations. Based on python packages of \texttt{GalSim} and \texttt{PythonPhot}, considering AT2017gfo-like kilonova fitted and generated by \texttt{MOSFiT}, we generate the simulated reference and science image and test this process with a flat background. We find that under the dark night-sky background of the site in Table \ref{tab:teles}, the detection depth considering the image subtraction is less than that calculated by comparing the kilonova magnitude with the limiting magnitude of the telescope. Given the spectrum of AT2017gfo-like kilonova, g and r bands are more effective in searching kilonova, which is consistent with the result in \cite{zhu_kilonova_2021} and \cite{chase_kilonova_2022}. If the kilonova is not detected around peak, the longer band can be useful to keep searching according to the spectrum evolution of kilonovae \citep{kasen_origin_2017}.
		\par
		Based on the flat background simulation, we add the host galaxy from the synthetic catalog CosmoDC2 into the image to explore the effect of the host. Using the stellar mass as the indicator to quantify the BNS merger rate of the galaxy, we filter out the galaxies which are less likely to be the host. The distance from the galaxy center is sampled following the offset distribution observed for short GRBs. We obtained the detectability at different redshifts and find that although the BNS mergers are farther away from the host’s center compared with SNe and TDEs, the host’s background can significantly affect kilonova detection and cannot be ignored. After taking into account the host galaxy, for the peak of AT2017gfo-like kilonova, the detection depth reduces from 357 (338) Mpc to 246 (213) Mpc in the g(r) band. The detection at longer wavelengths is affected more severely by the presence of a host galaxy, because of the spectral shape of BNS hosts. Given these influences, a longer exposure time is recommended in ToO follow-up. 
		\par
		We also compare the detection efficiency of WFST, LSST and ZTF, and their exposure times are inversely proportional to FoV. With 90\% detection efficiency, LSST proves to be the most efficient and can reach $\sim500$ Mpc with 40 s exposures. Although the FoV of ZTF is much larger than WFST, with deeper detection depth, WFST can do better than ZTF for distant kilonovae. For upcoming O4 and O5, WFST can coordinate with LSST appropriately to cover the probable region in the northern hemisphere. As the kilonova evolves and becomes dim after the peak, the detection depth of WFST decreases drastically, therefore the effective observation window is within the first two nights. To estimate how realistic is the choice of AT2017gfo as template, we construct a kilonova sample based on \texttt{bns} model and the fitting mass distributions of BNS. AT2017gfo belongs to the more luminous case in the sample, and the magnitude difference between sample and AT2017gfo is -0.5 to 1 mag, which means that the detection range can further decrease considering the kilonova diversity.
	    \par
		According to the detectability results above, we optimize the exposure time for every night and estimate the average total time to cover a region of GW localization in a ToO observation. Considering two nights and four hours each night to observe, with two bands WFST can well deal with the case that the localization region is no more than $\sim1000 \deg^2$ at 300 Mpc. Based on the simulation of BNS detection in O4 \citep{abbott_prospects_2020}, WFST can search and find most kilonovae if the probable sky region is observable and is predicted to find $\sim 30$\% kilonovae of BNS mergers reported in O4. 
		\par
		We note that in O3 ZTF applied package \texttt{gwemopt} to design and generate observation sequence \citep{coughlin_optimizing_2018,almualla_dynamic_2020}. A similar package can be used for WFST to increase the observation efficiency combined with the chosen exposure method in this work. 
   	\begin{acknowledgments}
		We appreciate the helpful discussions in WFST science team.
		This work is supported by the National Key R\&D Program of China Grant No. 2021YFC2203102 and 2022YFC2200100, NSFC No. 12273035, the Fundamental Research Funds for the Central Universities under Grant No. WK2030000036 and WK3440000004, and Cyrus Chun Ying Tang Foundations.
	\end{acknowledgments}
    \vspace{5mm} 
    \facilities{WFST}

    \software{\texttt{MOSFiT}\citep{Guillochon_2018, Villar_2017}, \texttt{GalSim}\citep{ROWE2015121}, \texttt{PythonPhot}\citep{2015ascl.soft01010J}, \texttt{NumPy}\citep{harris2020array, van2011numpy}, \texttt{Astropy}\citep{robitaille2013astropy}, \texttt{SciPy}\citep{virtanen2020scipy}, \texttt{matplotlib}\citep{hunter2007matplotlib}}


    
    \vspace{50mm}
    \bibliographystyle{aasjournal}
    \bibliography{draft}
    

\end{document}